\documentclass[aip,pop,reprint]{revtex4-1}
\usepackage{graphicx}
\usepackage[breaklinks=true,colorlinks=true,linkcolor=blue,urlcolor=blue,citecolor=blue]{hyperref}

\newcommand{\vc}{\mathbf}

\newcommand{\vtei}{\bar{v}_{T_{eI}}}

\newsavebox{\myimage}
\newcommand\redsout{\bgroup\markoverwith{\textcolor{red}{\rule[0.5ex]{2pt}{0.4pt}}}\ULon}

\draft 

\begin{document}

\title{Theory and simulation of anode spots in low pressure plasmas}

\author{Brett Scheiner}
\email[]{brett-scheiner@uiowa.edu}
\affiliation{Department of Physics and Astronomy, University of Iowa, Iowa City, Iowa 52240, USA}

\author{Edward V. Barnat}
\affiliation{Sandia National Laboratories, Albuquerque, New Mexico 87185, USA}

\author{Scott D. Baalrud}
\affiliation{Department of Physics and Astronomy, University of Iowa, Iowa City, Iowa 52240, USA}

\author{Matthew M. Hopkins}
\affiliation{Sandia National Laboratories, Albuquerque, New Mexico 87185, USA}

\author{Benjamin T. Yee}
\affiliation{Sandia National Laboratories, Albuquerque, New Mexico 87185, USA}

\date{\today}

\begin{abstract} 

When electrodes are biased above the plasma potential, electrons accelerated through the associated electron sheath can dramatically increase the ionization rate of neutrals near the electrode surface. 
It has previously been observed that if the ionization rate is great enough, a double layer separates a luminous high-potential plasma attached to the electrode surface (called an anode spot or fireball) from the bulk plasma.  Here, results of the first 2D particle-in-cell simulations of anode spot formation are presented along with a theoretical model describing the formation process. 
It is found that ionization leads to the buildup of an ion-rich layer adjacent to the electrode, forming a narrow potential well near the electrode surface that traps electrons born from ionization. 
Anode spot onset occurs when a quasineutral region is established in the potential well and the density in this region becomes large enough to violate the steady-state Langmuir condition, which is a balance between electron and ion fluxes across the double layer. 
A model for steady-state properties of the anode spot is also presented, which predicts values for the anode spot size, double layer potential drop, and form of the sheath at the electrode by considering particle, power, and current balance. 
These predictions are found to be consistent with the presented simulation and previous experiments. 
\end{abstract}


\maketitle

\section{introduction}

\begin{figure}
\begin{center}
\includegraphics[width=8.5cm]{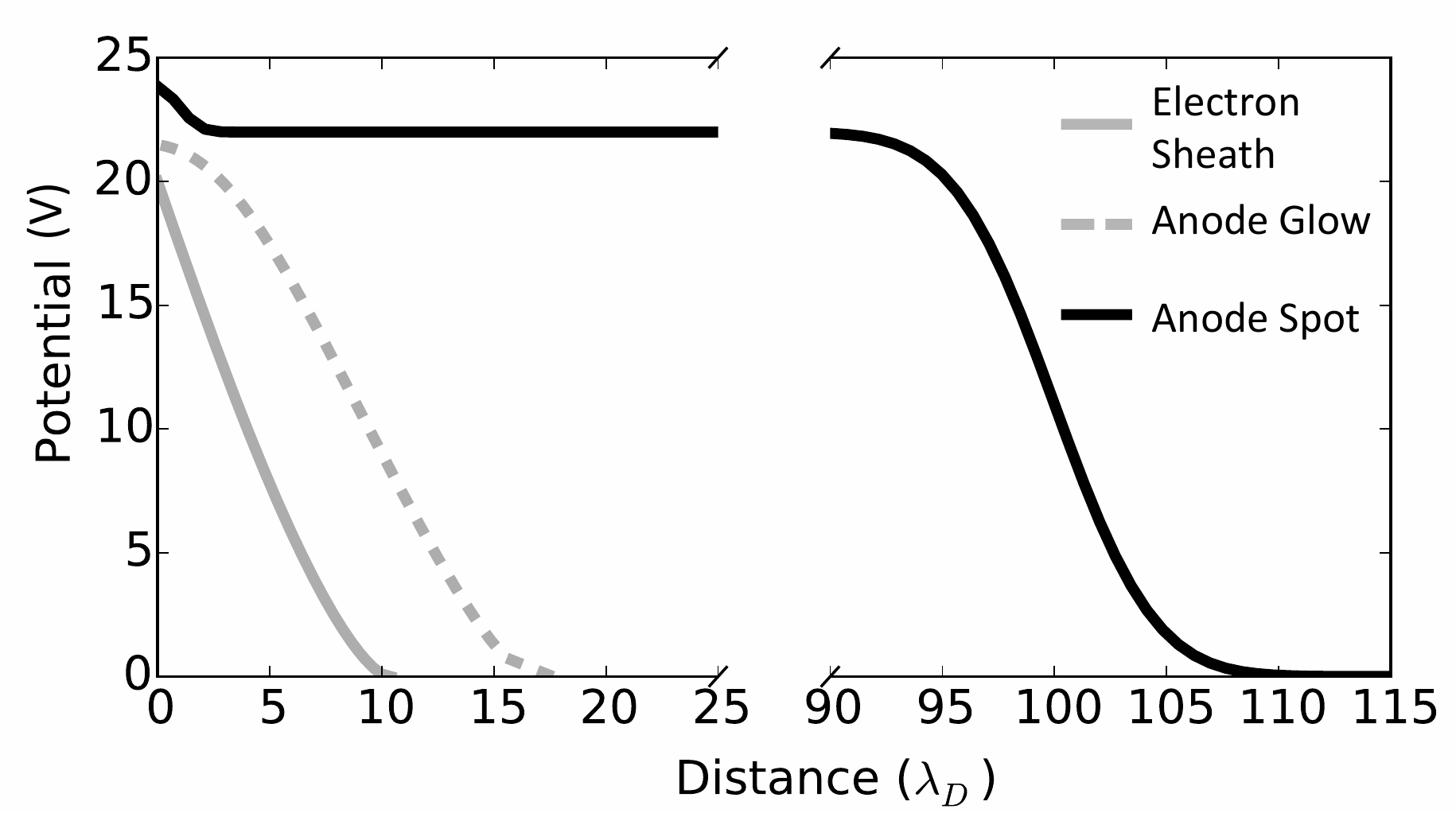}
\end{center}
\caption{Sketch of potential profiles for three example sheath-like structures that are possible near positively biased electrodes: An electron sheath (grey solid line), an anode glow (grey dashed line), and an anode spot (black solid line). \label{fg:sketch}}
\end{figure}

\begin{figure}
\begin{center}
\includegraphics[width=8.5cm]{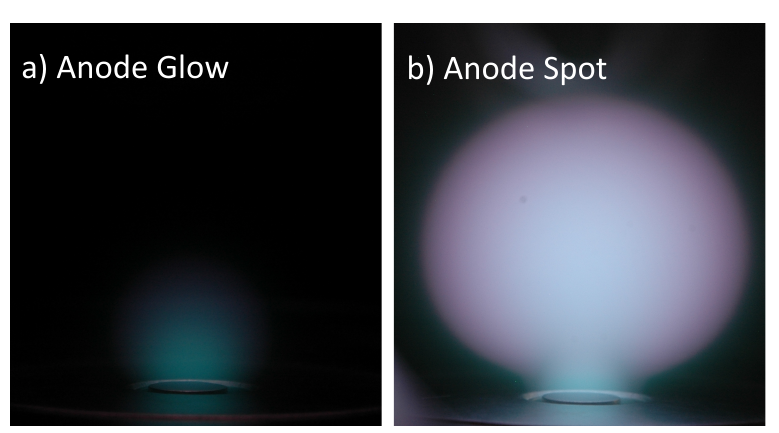}
\end{center}
\caption{ Photograph of (a) an anode glow and (b) an anode spot. \label{fg:img}
 }
\end{figure}

Anode spots are a discharge phenomenon that occur near positively biased electrodes. When the potential across an electron sheath is large, electrons can gain enough energy to ionize neutral species by electron impact ionization. If the ionization rate is sufficient, a positive space charge layer will develop at the electrode next to a negative space charge layer adjacent to the plasma. These adjacent positive and negative space charge layers are a type of anode double layer \footnote{Note that the term {\emph{double layer}} usually refers to a similar structure between a high and low potential plasma with equal amounts of positive and negative charge, hence the need to distinguish anode double layers.} called an anode glow \cite{2009PSST...18c5002B}. The anode glow potential profile is sketched in Fig.~\ref{fg:sketch} and is shown in the photograph in Fig.~\ref{fg:img}a. In experiments with sufficient ionization, it has been observed that an increase in bias leads to the rapid formation of an anode spot\cite{1929PhRv...33..954L,1979JPhD...12..717T,2008PSST...17c5006S,2009PSST...18c5002B}. The main characteristic of an anode spot is a double layer detached from the electrode separating a luminous high-potential plasma from the bulk plasma. An anode spot is shown in the photograph in Fig.~\ref{fg:img}b. Potential profiles typical of these situations are shown in Fig.~\ref{fg:sketch}. Anode spots are just one of the several forms of sheath structure that are possible near positivly biased electrodes\cite{1929PhRv...33..954L} and are particularly of interest due to their application in electron temperature control\cite{2013PSST...22f5002Y}, dust confinement\cite{1995PhPl....2.3261B}, plasma contactors\cite{1996PhPl....3.3875A}, as an ion source \cite{2011RScI...82l3303P}, and as a platform to study plasma self organization \cite{2013JMPh....4..364S,2011ITPS...39.2118M}. In this paper, the anode spot onset is studied in particle-in-cell simulations for the first time and a theory for their onset and equilibrium properties is presented.

Stationary double layers are expected to satisfy a flux density balance criterion called the Langmuir condition\cite{1978Ap&SS..55...59B},
\begin{equation}\label{lc}
 \Gamma_i=\sqrt{\frac{m_e}{m_i}}\Gamma_e,
\end{equation}
where $\Gamma_e=n_eV_e$ and $\Gamma_i=n_iV_i$ are the electron and ion flux densities and $V_e$ and $V_i$ are the electron and ion flow velocities at the double layer sheath edge.  In 1972 Block\cite{1972CosEl...3..349B} noted that the Langmuir condition is satisfied in the reference frame of the double layer. Two decades later Song, Merlino, and D'Angelo \cite{1992PhyS...45..391S} derived a modified Langmuir condition by considering the momentum and continuity equation in the frame of a double layer moving with velocity $U_{DL}$ with respect to the lab frame. The condition in the lab frame is 
\begin{equation}\label{ml}
 \Gamma_i-n_{Hi}U_{DL}=\sqrt{\frac{m_e}{m_i}}\Gamma_e,
\end{equation}
where $n_{Hi}$ is the density of the plasma at the double layer's high potential side. Here, the double layer motion results from an imbalance in the electron and ion fluxes crossing the double layer. Using this newly derived condition, the stability of the anode spot steady-state was studied\cite{1992JPhD...25..938S}. 

In this paper, the modified Langmuir condition is used to derive a density criterion between the high and low potential sides of the double layer required for spot onset. This density criterion is then used to form a model for the critical electrode bias required for spot onset. When a flux imbalance in Eq.~(\ref{lc}) is present, the double layer is predicted to move away from the electrode forming an anode spot. The model, informed by particle-in-cell (PIC) simulations, also describes a mechanism for establishing a quasineutral region with $n_i\approx n_e$ and $E\approx 0$ at the high potential side. Once sufficient ion density has built up next to the electrode, a potential well for electrons forms. This allows trapping of electrons near the electrode and the formation of a quasineutral spot plasma. Previous work had suggested that an instability was needed for spot onset.\cite{1981JPhD...14.1403A} The model presented here suggests a significantly different picture that does not include instability as a feature of spot onset. 

This paper also presents a model for steady-state properties of the anode spot. In the model, the balance of particle creation and loss, power, and current loss within the spot are used to determine the anode spot diameter, double layer potential drop, and form of the sheath between the electrode and spot plasma as a function of electron temperature. This model couples these different properties through the energy dependence of the electron impact ionization cross section of neutrals. Prior descriptions of the anode spot length scale $L$ have used equations of the form  
 \begin{equation}\label{ls}
 \sqrt{\frac{m_e}{m_i}}L\sigma_I  n_n \approx 1,
 \end{equation}
where $\sigma_I$ is the electron impact ionization cross section and $n_n$ is the neutral density\cite{1991JPhD...24.1789S,2009PSST...18c5002B}. In these descriptions, a constant estimate of the cross section was assumed instead of considering its energy dependence.

This paper presents the first computational study of anode spots. The simulations presented in Sec.~\ref{sec:sim}, are used to formulate a theory for the spot onset in Sec.~\ref{sec:onset}. Following this, the theory of Sec.~\ref{sec:current} and Sec.~\ref{sec:power} predicts steady-state properties of the anode spot. Finally, properties of the ion presheath leading up to the high potential entrance of the double layer are calculated in Sec.~\ref{sec:presheath}, showing that under typical low temperature plasma conditions the presheath length is determined by the electron impact ionization rate. Concluding remarks are given in Sec.~\ref{sec:conclusion}. 

\section{SIMULATIONS\label{sec:sim}}
\subsection{Simulation setup}

Simulations were performed using the electrostatic PIC code Aleph \cite{2012CoPP...52..295T}.  Aleph solves for the electric field on an unstructured mesh, and includes three velocity components for particles. The simulations incorporated collisions between charged and neutral particles using the Direct Simulation Monte-Carlo (DSMC) method\cite{bird1998molecular}.

\begin{figure}
\begin{center}
\includegraphics[width=8cm]{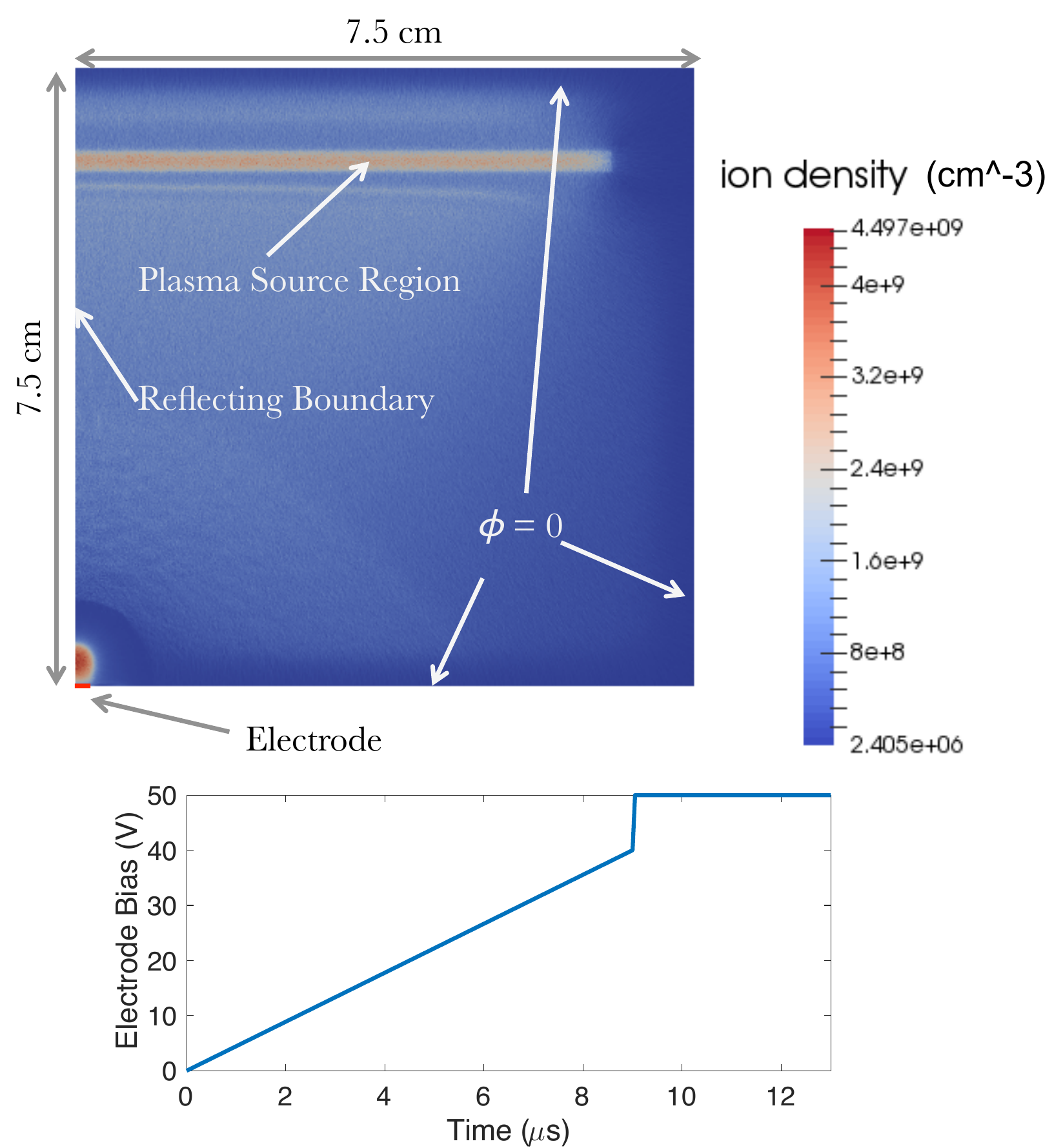}
\end{center}
\caption{The 7.5 cm by 7.5 cm simulation domain with the color map indicating a typical example of the ion densities encountered in the simulations. An anode spot is attached to the electrode in the lower left hand corner of the domain. \label{fg:setup} }
\end{figure}

The 2D simulation domain shown in Fig.~\ref{fg:setup} was chosen to resemble experiments\cite{2014PhPl...21j3512B}, though at reduced spatial dimensions of 7.5 cm by 7.5 cm. The boundary conditions for particles were one reflecting boundary (left wall) and three absorbing boundaries (right, top, and bottom walls). The absorbing walls had a $\phi=0$ V Dirichlet boundary condition for the electric field except for a small 0.25 cm electrode embedded in the lower boundary. The electrode bias was increased linearly from 0 V at t = 0 to 40 V at t = 9 $\mu s$. At $t=9 \mu$s, the electrode bias was stepped to a fixed value and held constant for the remainder of the simulation; see Fig.~\ref{fg:setup}. The final electrode bias was varied over several simulations to determine that required for anode spot formation. The reflecting boundary had a Neumann $\nabla \phi\cdot\hat{n}=0$ boundary condition, which in combination with the reflecting boundary condition for particles allows the simulation to represent half of a symmetric 15 cm by 7.5 cm domain. The domain was discretized using an unstructured triangular mesh with an element size of approximately 0.0116 cm that resolved the Debye length across the domain throughout the duration of the simulation. A time step of 10 ps was chosen to resolve the CFL condition for all particles, preventing particles from crossing more than one cell length before a velocity update and ensuring that particle trajectories were resolved.

A helium plasma was continuously generated at a rate of $4.7\times 10^9 \ \textrm{cm}^{-3} \mu s^{-1}$ in a 0.25 cm by 6.5 cm source region 6.25 cm above the electrode. This was designed to model a source chamber in the experiments described in Ref.~\onlinecite{2014PhPl...21j3512B}.  Electrons and ions were sourced with temperatures of 4 eV and 86 meV respectively. Particles left the source region by ambipolar diffusion and filled the domain. After expansion of the plasma from the source region, the electron temperature in the bulk plasma was approximately 2.4~eV.  A neutral helium background with density $7\times10^{15} \ \text{cm}^{-3}$ at 24 meV (200 mTorr at 273 K) was present in the simulation. Electrons, ions, and neutral particles had weights of 2000, 8000, and $2\times10^{10}$ respectively. Two populations of electrons were tracked to better understand the effects of ionization within the anode spot. Electrons which have participated in an ionization event are denoted $e^-_I$ and those which have not are denoted $e^-_B$ (bulk electrons). The included ionization interactions for these species are
\begin{equation}\label{ei1}
e^-_{B}+ \textrm{He} \to 2 e^-_I +\textrm{He}^+
\end{equation}
and
\begin{equation}\label{ei2}
e^-_{I}+ \textrm{He} \to 2 e^-_I +\textrm{He}^+.
\end{equation}
Elastic electron neutral collisions with the helium background were also included. The cross sections used for collisions in the simulations and for other calculations in the following sections were obtained from the Phelps database on LXcat\cite{lxcat}.

\subsection{Anode spot simulation\label{sec:onsetsim}}

Simulations were carried out for final electrode biases of 40 V, 45 V, 46 V, 47 V, 48 V, and 50 V to determine the critical electrode bias for spot onset. For each simulation, the plasma potential at t = 9 $\mu s$, just before the electrode bias was increased, was 6.4 V. Anode spots were observed for the simulation with a 48 V electrode bias, but not with a 47 V bias, indicating that the critical bias with respect to the plasma potential is in the range 40.6-41.6 V. In this section, the simulation with a 50 V bias will be used to study the spot onset. 

  \begin{figure}
  \includegraphics[width=8.5cm]{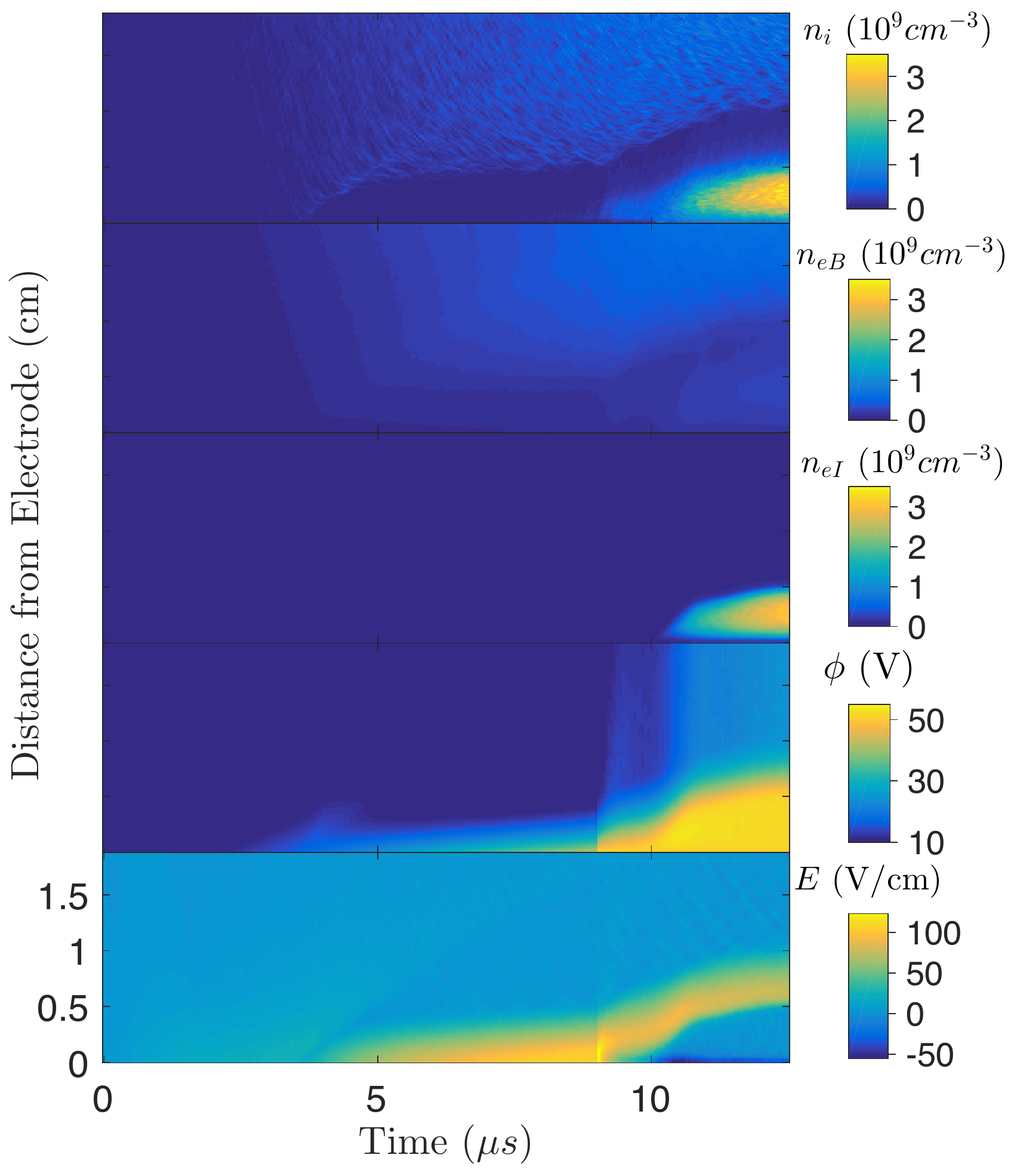}
   \caption{Time-dependent values of plasma species densities, potential, and electric field along the symmetry axis (reflecting boundary). Spot onset follows the increase in electrode bias at $t=9 \ \mu s$. \label{fg:t_profiles}}
 \end{figure}

Figure~\ref{fg:t_profiles} shows profiles of the ion, bulk electron, and ionization electron densities along with the electrostatic potential and electric field along the symmetry axis (reflecting boundary) in time throughout the 50 V simulation. 
Early in the simulation, particles were observed to fill the domain by ambipolar diffusion from the source region. Initially, a large portion of the lower half of the domain was electron rich. After approximately $4 \ \mu s$, the ions had enough time to traverse the domain and encounter the electric field at the electrode. Once these ions were reflected by the field, an electron sheath formed. This can be seen in the ion density and electric field in Fig.~\ref{fg:t_profiles}. 
Between $5 \ \mu s$ and $9 \ \mu s$, waves propagating towards the electron sheath can be seen in the ion density and electric field. These are ion acoustic waves excited by the differential flow between ions in the plasma and electrons accelerating to their thermal speed in the electron presheath \cite{2015PhPl...22l3520S}.

 \begin{figure*}
 \includegraphics[width=.9\textwidth]{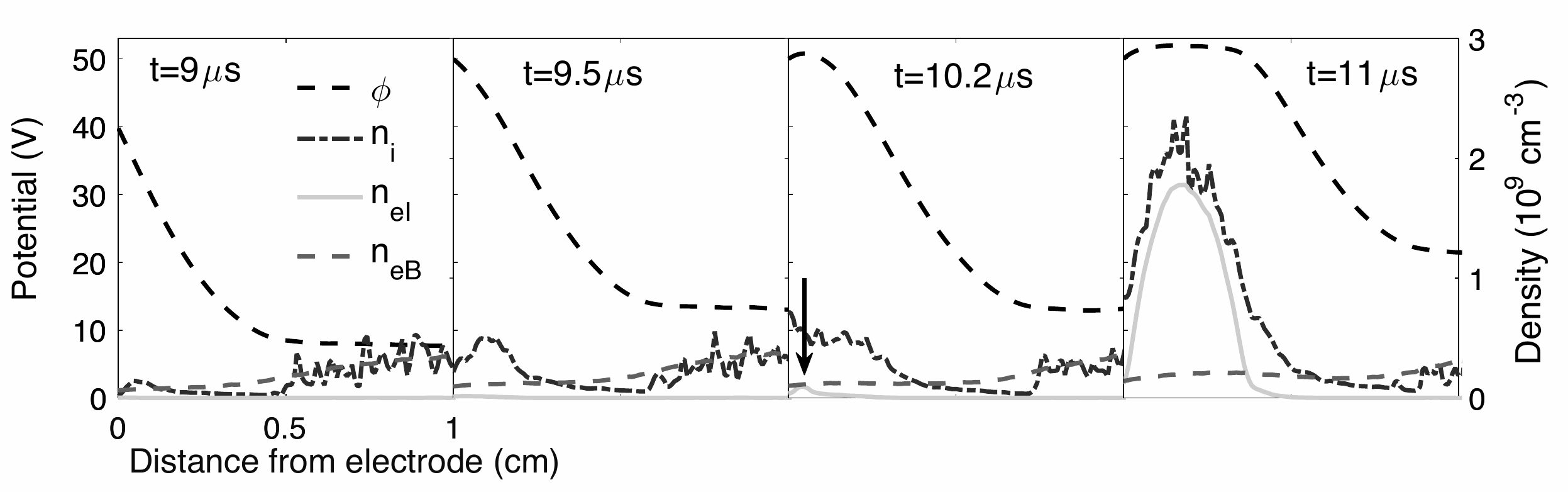}
 \caption{The potential and density profiles just before ($t=9 \ \mu s$) and after the electrode bias increases from 40 V to 50 V. Note that after the potential gradient changes sign around $t=10.2 \ \mu s$ the density of electrons from ionization begins to increase, this feature is highlighted by the black arrow.  \label{fg:simp}
 }
 \end{figure*}

At $9 \ \mu s$ the electrode bias was set to 50 V. Figure~\ref{fg:t_profiles} shows a dramatic increase in ion density within the sheath immediately following this jump in applied bias. The effect of this increase can be seen by comparing the first two panels of Fig.~\ref{fg:simp}. The increase in ion density is not immediately accompanied by an increase in $n_{eI}$ since the sheath field quickly accelerates ionization-born electrons to the electrode. At $t=9.5 \ \mu s$, the buildup of ions within the sheath has resulted in a change in concavity of the potential profile. The panel of Fig.~\ref{fg:simp} marked $t=10.2 \ \mu s$ shows that a further increase in ion density results in a change in sign of the electric field, which is indicated by the maximum in the potential just off of the electrode surface accompanied by a small increase in $n_{eI}$. Low energy electrons resulting from electron impact ionization interactions have temperature of approximately 1 eV and  are trapped at the maximum by the $\sim1$ V drop to the electrode. Once this electron trapping occurs, the electron density increases rapidly until quasineutrality is established. This is followed by expansion leading to the formation of the anode spot plasma shown in the densities and electric field plotted in Fig.~\ref{fg:t_profiles} and the panel marked $t=11 \ \mu s$ in Fig.~\ref{fg:simp}. 

By t = 13 $\mu s$, expansion of the anode spot slows and the double layer potential settles at approximately $27.2$ V, which is 2.6 V above the potential needed for electron impact ionization of helium. This is similar to experiments which have measured the double layer potential to be a few volts above the ionization energy of the neutral gas\cite{1991JPhD...24.1789S,1987PhFl...30.2549C,2007PhPl...14d2109B}. The plasma potential within the spot settles at around 51 V, which is 1 V above the electrode potential. The scale length of the spot at this time ($13 \ \mu s$) is 0.5 cm, indicated by the 2D geometry of the zero electric field area shown in Fig.~\ref{fg:2d}.


 
 
 \begin{figure}
\includegraphics[width=8cm]{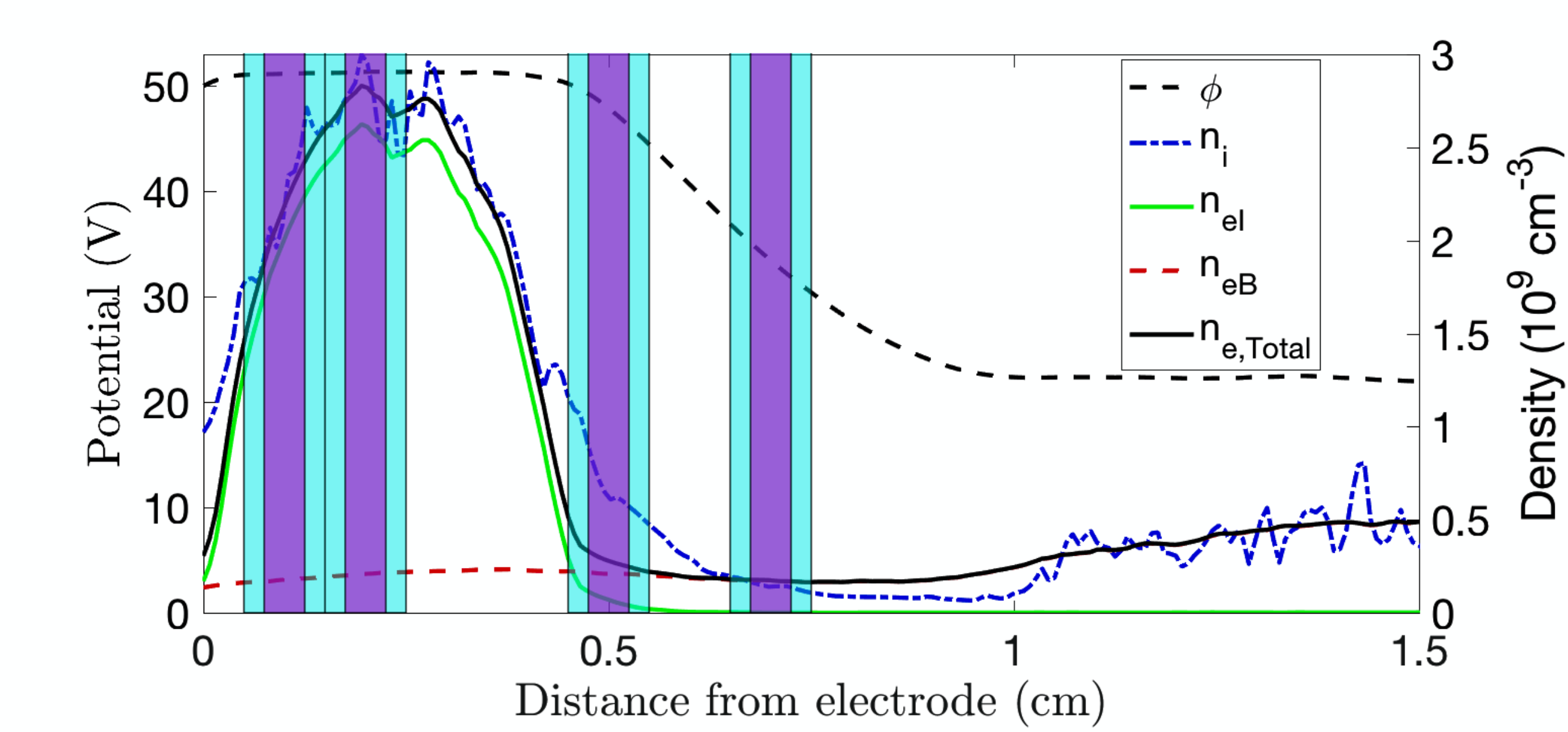}
 \caption{ Particle species density and potential profiles at $t=10.8\mu s$. The narrower purple shaded regions indicate the area in which ion VDF histograms were obtained for the data shown in Fig.~\ref{fg:vdf}, and wider blue shaded regions show the same for electrons. \label{fg:vdfloc} }
 \end{figure}

   \begin{figure*}
\includegraphics[width=15cm]{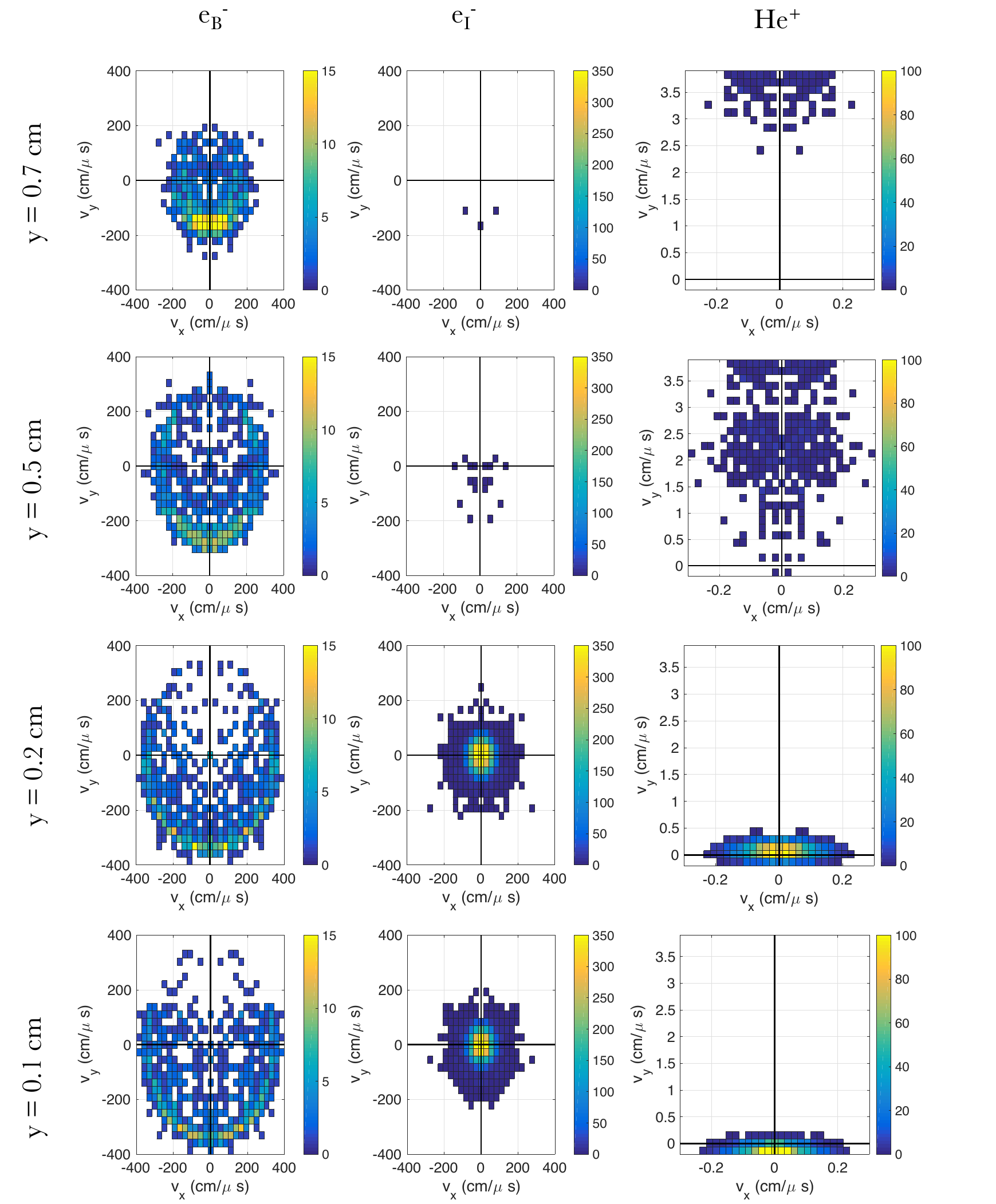}
 \caption{Velocity distribution functions for $e^-_B$ and $e^-_I$ electrons and helium ions within the anode spot and double layer. The color bars indicate the number of macroparticles per bin. These VDFs were obtained at t=10.8 $\mu s$ of the simulation at the locations indicated in Fig.~\ref{fg:vdfloc}. Note that the coordinate axes for ions span different ranges in $v_x$ and $v_y$.  Also note that the area in which ion histograms were obtained was one quarter of that used for electrons; see Fig.~\ref{fg:vdfloc}.\label{fg:vdf} }
 \end{figure*}


\subsection{Velocity distribution functions}


The VDFs were examined at $t=10.8 \ \mu s$ by taking histograms of particle velocities in four 2D regions inside the anode spot. Ion data was obtained in a region 0.05 cm by 0.05 cm, and electron data in a 0.1 cm by 0.1~cm region; see Fig.~\ref{fg:vdfloc}. Different sized regions were used due to the different macroparticle weight of electrons and ions. 
The corresponding VDFs are shown in Fig.~\ref{fg:vdf}. Section II of Ref. \onlinecite{2016PhPl...23h3510S} provides further details regarding the calculation of VDFs along the reflecting boundary.

Figure~\ref{fg:vdfloc} shows that at $y=0.7$ cm, a location within the double layer, $e^-_B$ electrons have a peak density near $v_y=-180$ cm/$\mu s$ corresponding to the $\approx10$ eV energy gained from the double layer electric field between 1 cm and 0.7 cm. These electrons increase in kinetic energy to nearly 26 eV at the maximum potential near $y=0.1-0.2$ cm. This corresponds to the half-ring-like distribution with a peak density near $v\approx300$ cm/$\mu$s. The geometry of the VDF within the spot is due to the curvature of the surface of the double layer and the multitude of directions from which electrons enter the spot. Electrons born from ionization ($e^-_I$) within the spot have a nearly Maxwellian VDF. Few of these electrons are present between y = 0.5-0.7 cm since they are trapped by the strong double layer electric field. The VDFs also show that this population is much more dense than that of electrons from the bulk. This can also be seen in the density profiles of Fig.~\ref{fg:vdfloc}.

Ions born within the spot are well described by a Maxwellian distribution with no flow shift at $y = 0.2$~cm. However, very close to the electrode ($y=0.1$~cm) ions are flow shifted due to the ion presheath leading up to the electrode surface. Within the double layer, ions are accelerated towards the bulk plasma as indicated by their velocity and decreased density.  

\subsection{Current collection}

\begin{figure}
\includegraphics[width=8.5cm]{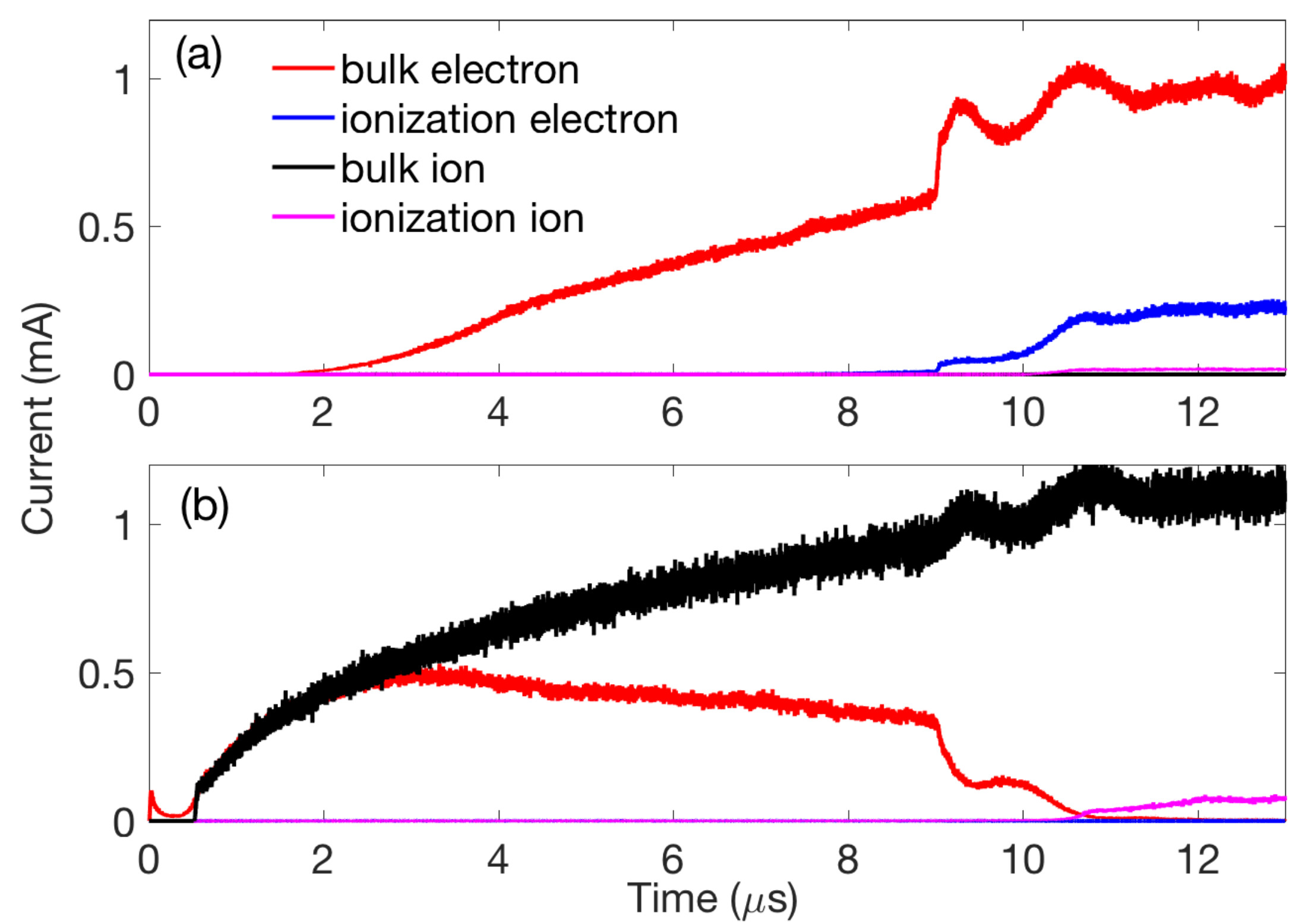}
 \caption{Components of the current collected by the electrode (a) and wall (b). After anode spot onset the electron current collection increases. When the spot is present at $t>10\mu$s, all electrons lost from the simulation exit through the electrode. \label{fg:current}  }
\end{figure}

One common feature observed in experiments is a jump in the current collected by the electrode when a spot forms~\cite{2009PSST...18c5002B,1991JPhD...24.1789S,1987PhFl...30.2549C}. It has previously been suggested that the cause of this increase is an increase of electron collecting area resulting from the relatively large surface area of the double layer compared to the electrode\cite{2008PSST...17c5006S}. 
Figure \ref{fg:current} shows that a current increase was also observed in the simulations. 
Most of the current collected at the electrode is due to the bulk plasma electrons ($e^-_B$) accelerated to the electrode by the double layer. There is also a $\approx 20\%$ contribution from electrons formed by ionization after spot onset. This confirms that the majority of the increased current collection after onset is due to the increased collection area for electrons from the bulk plasma. In Fig.~\ref{fg:current}, the current component due to ions born from ionization collected at the walls of the domain is also shown. This indicates that approximately 5-10\% of the ion current collected at the walls after spot onset is due to ions born in the spot and accelerated to the bulk plasma by the double layer. 
Once the anode spot has formed, all of the bulk ion current is collected by the walls and all the bulk electron current is lost to the electrode. This situation, known as global non-ambipolar flow\cite{2007PhPl...14d2109B}, occurs near electrodes that are small enough to be biased above the plasma potential and large enough to collect the entirety of the electron current. 
In this case, it is the effective collection area of the anode spot that establishes this flow scenario. 
After onset, the current shows oscillations on a microsecond timescale, possibly related to the increased fluctuations in the ion density and electric field shown in Fig.~\ref{fg:t_profiles}.

\begin{figure}
\includegraphics[scale=0.32]{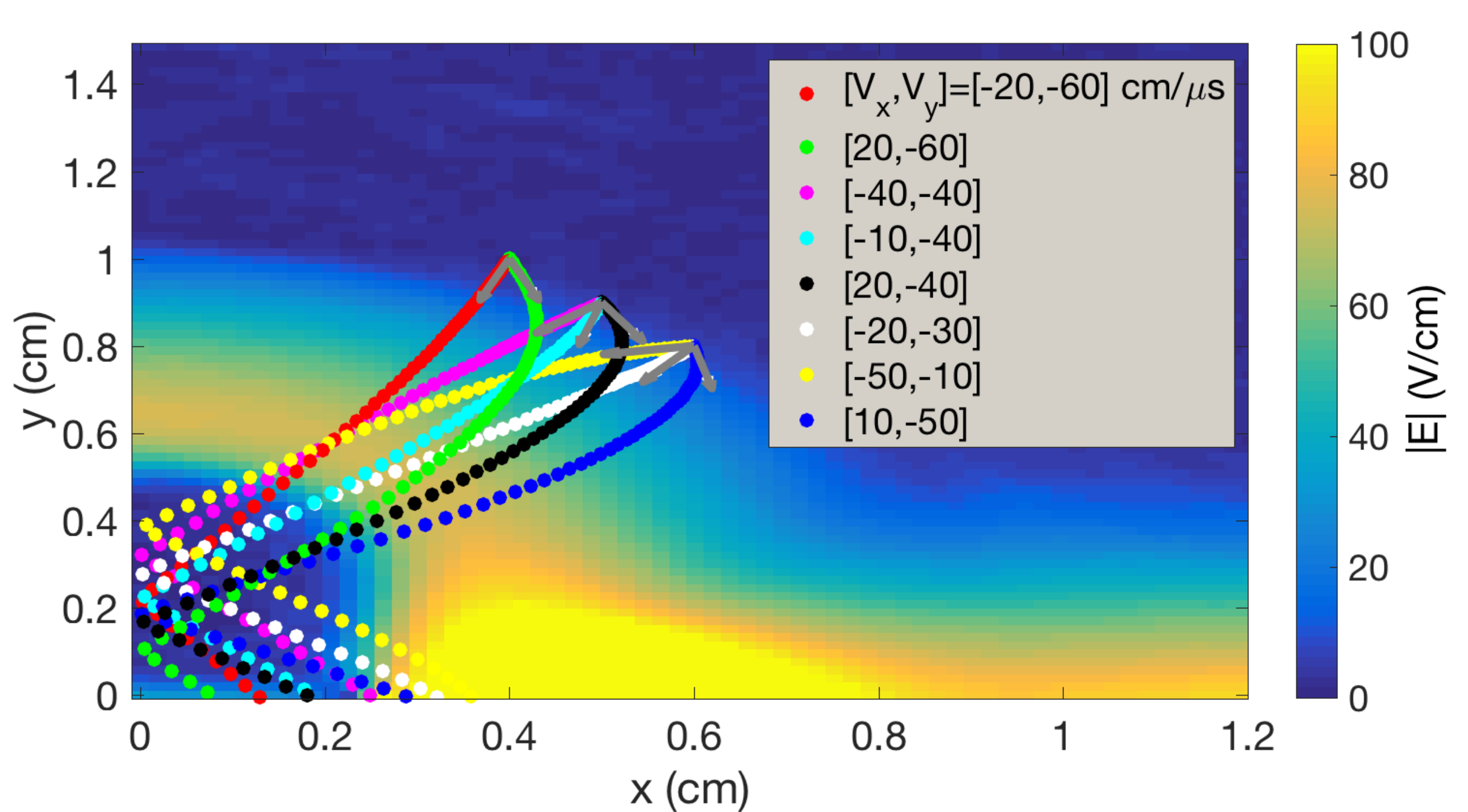}
 \caption{Test electron trajectories integrated in the electric field obtained from the 50 V simulation of Sec.~\ref{sec:onsetsim}. The initial velocity vectors are shown in the legend and are representative of electron velocities found near the sheath edge. Electrons starting in the range $x\approx[0,0.4]$ are collected by the electrode. \label{fg:2d}}
\end{figure}

Additional evidence for the increase in electron current collection area can be found by integrating test electron trajectories starting at locations outside the double layer. The integration of the trajectories used the velocity Verlet method along with the electric field from a single time step of the simulation at $t=10.8 \ \mu s$. The electric field within the anode spot had small fluctuations on the timescale of an electron crossing the spot, but these contributed a negligible change in the trajectories so the time dependence of the electric field was neglected. Selected particle trajectories are plotted over the electric field in Fig.~\ref{fg:2d}. The initial velocities for the particles are typical of those found in a flow-shifted distribution at the sheath edge. Trajectories starting between $x=0$ cm and $x\approx0.4$ cm impact the electrode, which extends from 0 to 0.25 cm on the lower boundary. Particles starting further to the right typically impact other parts of the lower boundary. These trajectories show that the anode spot increases the current collection area. 

\section{THEORY}

\subsection{Anode spot onset\label{sec:onset}}
 
Section~\ref{sec:onsetsim} showed simulation results demonstrating that anode spot onset is a consequence of the buildup of positive space-charge within the sheath due to electron impact ionization of neutral atoms. Initially, electrons born from ionization were quickly accelerated to the electrode by the sheath electric field, but when the positive space-charge grew large enough, the potential profile developed a local maximum off the electrode surface. This local maximum is a potential well for electrons. Low energy electrons born from ionization within this well are trapped, allowing the electron density to increase. It was observed that this trapping led to the formation of a quasineutral region where E $\approx$ 0 and $n_e\approx n_i$.

Once quasineutrality is established, the double layer is predicted to move if there is an imbalance in the Langmuir condition, resulting in the double layer velocity given in Eq.~(\ref{ml}). To determine when the double layer will move outward, resulting in the expansion of the anode spot plasma, the flux densities entering the double layer are considered. Bulk electrons are accelerated to their thermal speed ($v_{T_e}=\sqrt{T_e/m_e}$) in an electron presheath \cite{2015PhPl...22l3520S,2017PSST...26b5009Y} leading up to the double layer at the low potential side. This flux density is 
\begin{equation}\label{gme}
\Gamma_e=n_{eB}\sqrt{\frac{T_{e,B}}{m_e}},
\end{equation} 
where the subscript B indicates values in the bulk plasma. Ions enter the high potential side of the double layer near the electrode at their sound speed due to the Bohm criterion, resulting in the flux density
\begin{equation}\label{gmi}
\Gamma_i=n_{i,Hi}\sqrt{\frac{T_{e,Hi}}{m_i}},
\end{equation}   
where the subscript Hi indicates values at the high potential side of the double layer. Inserting Eqs.~(\ref{gme}) and (\ref{gmi}) into the modified Langmuir condition Eq.~(\ref{ml}) and requiring $U_{DL}>0$ as a condition for anode spot onset\footnote{Here, onset is defined as the expansion of the quasineutral region. This coincides with the motion of the double layer.} leads to 
\begin{equation}\label{on}
n_{i,Hi}>n_{eB}\sqrt{\frac{T_{e,B}}{T_{e,Hi}}}.
\end{equation}
Prior to formation of the potential well, it may be thought that the modified Langmuir condition, Eq.~(\ref{ml}), suggests $U_{DL}<0$ so the double layer will move towards the electrode. However, Eq.~(\ref{ml}) is only valid when there is a region with E $\approx$ 0 at both the high and low potential sides of the double layer. Thus, quasineutrality must first be established on the high potential side, creating a region with $E \approx 0$, before an imbalance in the static Langmuir condition implies expansion of the double layer. 

Previous attempts to model spot onset solved an integral form of Poisson's equation\cite{2006PhPl...13k3504C,1981JPhD...14.1403A,1983JPhD...16..601A}. In these models, the integral of the charge density within the sheath, including contributions from electron impact ionization, were formulated as a function of the sheath electric field   
\begin{equation}\label{p1}
\frac{1}{2}(E)^2_{\phi=\phi_o}=4\pi e \int_0^{\phi_o} \rho (E(\phi^\prime)) d\phi^\prime.
\end{equation}
Models based on Eq.~(\ref{p1}) do not provide solutions which include a potential well for electrons because it requires that the electric field becomes multivalued as a function of the potential.  

For typical experimental conditions where $T_{e,B}/T_{e,Hi}\simeq1$, Eq.~(\ref{on}) predicts that the double layer will move when the high potential density  exceeds the bulk density. Fig.~\ref{fg:t_profiles} shows that when this occurs the double layer begins to move outward, which is in agreement with the condition in Eq.~(\ref{on}). After 10.7 $\mu s$ the double layer position indicated by the electric field in Fig.~\ref{fg:t_profiles} shows that its speed has decreased. At first glance this may seem to contradict Eq.~(\ref{ml}) since the peak density within the center of spot has increased significantly even though the double layer velocity has not. However, a careful inspection of the density and potential profiles at $t \ = \ 10.8 \ \mu s$ in Fig.~\ref{fg:vdfloc} reveals that the plasma density at the high potential sheath edge is slightly less than half the peak density within the spot. The high potential density was still large enough for the double layer to continue to slowly move outward. The theory predicts that when the densities at the entrance to the double layer balance the Langmuir condition its motion stops. The simulation time was not long enough for a steady-state configuration to be observed.    

To predict when an anode spot will form, it is desirable to connect the density criterion of Eq.~(\ref{on}) with the electrode bias. In experiments, anode spot onset is observed when the electrode exceeds some critical bias. The critical bias can be predicted by considering the conditions required for spot onset. Two conditions are required: 1) a quasineutral region is established at the high potential side of the double layer so that the assumptions of the modified Langmuir condition are satisfied and the double layer is free to move, and 2) the ion flux leaving the high potential side leads to an imbalance in the Langmuir condition resulting in motion of the double layer and expansion of the anode spot plasma. Electrons born from ionization near the maximum of the potential profile are trapped if their energy is less than the depth of the  well. A potential well for electrons is a potential hill for ions, therefore, for a deep enough well, the electron density increases faster than the ion density. This occurs if the depth of the well is greater than the magnitude of the floating potential. Once the electron density increases to match the ion density at the bottom of the well condition 1) is satisfied. If ions are born faster than they are lost, the plasma density within the electron potential well will increase until condition 2) is satisfied. The statement that the ion birth rate is greater than the ion loss rate can be expressed as
\begin{equation}\label{cb1}
\underbrace{2A_{\text{sheath}}n_{Hi}c_{s,Hi}}_{\textrm{loss rate}}<\underbrace{n_n\Gamma_e\sigma_IA_{\text{sheath}}z_{\text{sheath}}}_{\textrm{ionization rate}},
\end{equation}
where $A_{\text{sheath}}$ is the cross-sectional area of the sheath prior to the double layer expansion, $c_{s,Hi}$ is the ion sound speed at the high potential side, and $z_{\text{sheath}}$ is the sheath thickness. The numerical factor is due to half of the ions born near the electron well being lost in the electrode direction and the other half being lost in the plasma direction.

\begin{figure}
\includegraphics[scale=0.35]{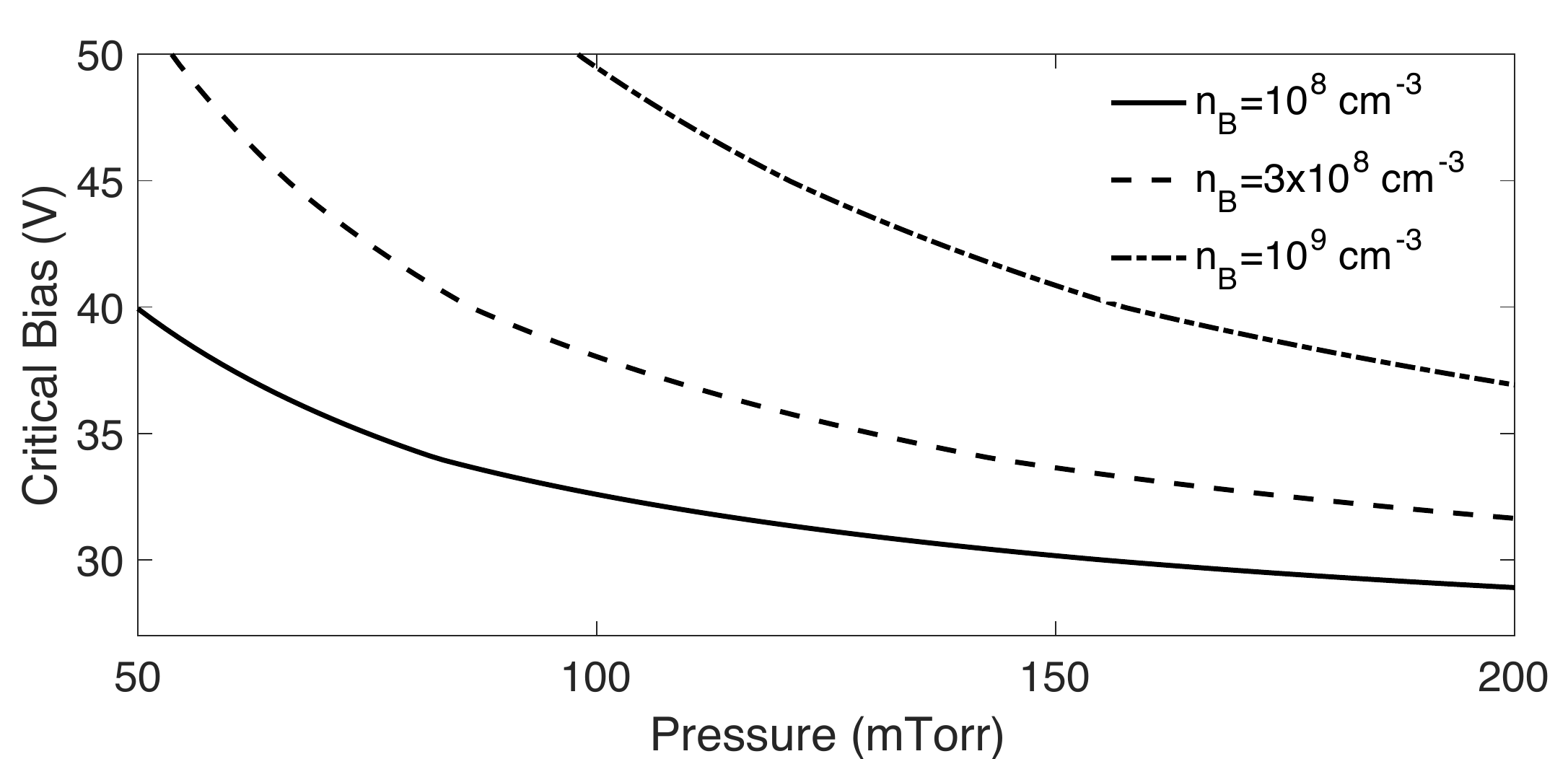}
\caption{The critical bias as a function of pressure for several different bulk plasma densities and an electron temperature of 2.4eV calculated from Eq. (\ref{tc}) with the sheath thickness of Eq. (\ref{esh}).\label{fg:cb1} }
\end{figure}

To determine the critical bias, the condition of exact balance in Eq. (\ref{cb1}) is considered. Combining this with the steady-state Langmuir condition of Eq. (\ref{lc}) and the high potential ion flux in Eq. (\ref{gmi}), the threshold condition becomes
\begin{equation}\label{tc}
1-\frac{1}{2}\sqrt{\frac{m_i}{m_e}}n_n\sigma_Iz_{\text{sheath}}=0.
\end{equation}
This is similar to the equation used by Song et al.\cite{1991JPhD...24.1789S} to estimate the sheath thickness of the anode glow prior to spot onset. 
Using this condition along with an independent estimate of the electron sheath thickness\cite{2015PhPl...22l3520S} 
\begin{equation}\label{esh}
z_{\text{sheath}}=0.79\lambda_{De}\bigg(\frac{e\Delta\phi}{T_e}\bigg)^{3/4} ,
\end{equation}
Eq. (\ref{tc}) can be solved numerically for the critical bias $\Delta\phi_c$ given the energy-dependent electron impact ionization cross section $\sigma_I(e\Delta\phi_c)$. The critical bias for a helium plasma with an electron temperature of 2.4~eV is shown in Fig.~\ref{fg:cb1} for several values of the bulk plasma density.

\begin{figure}
\includegraphics[width=8.5cm]{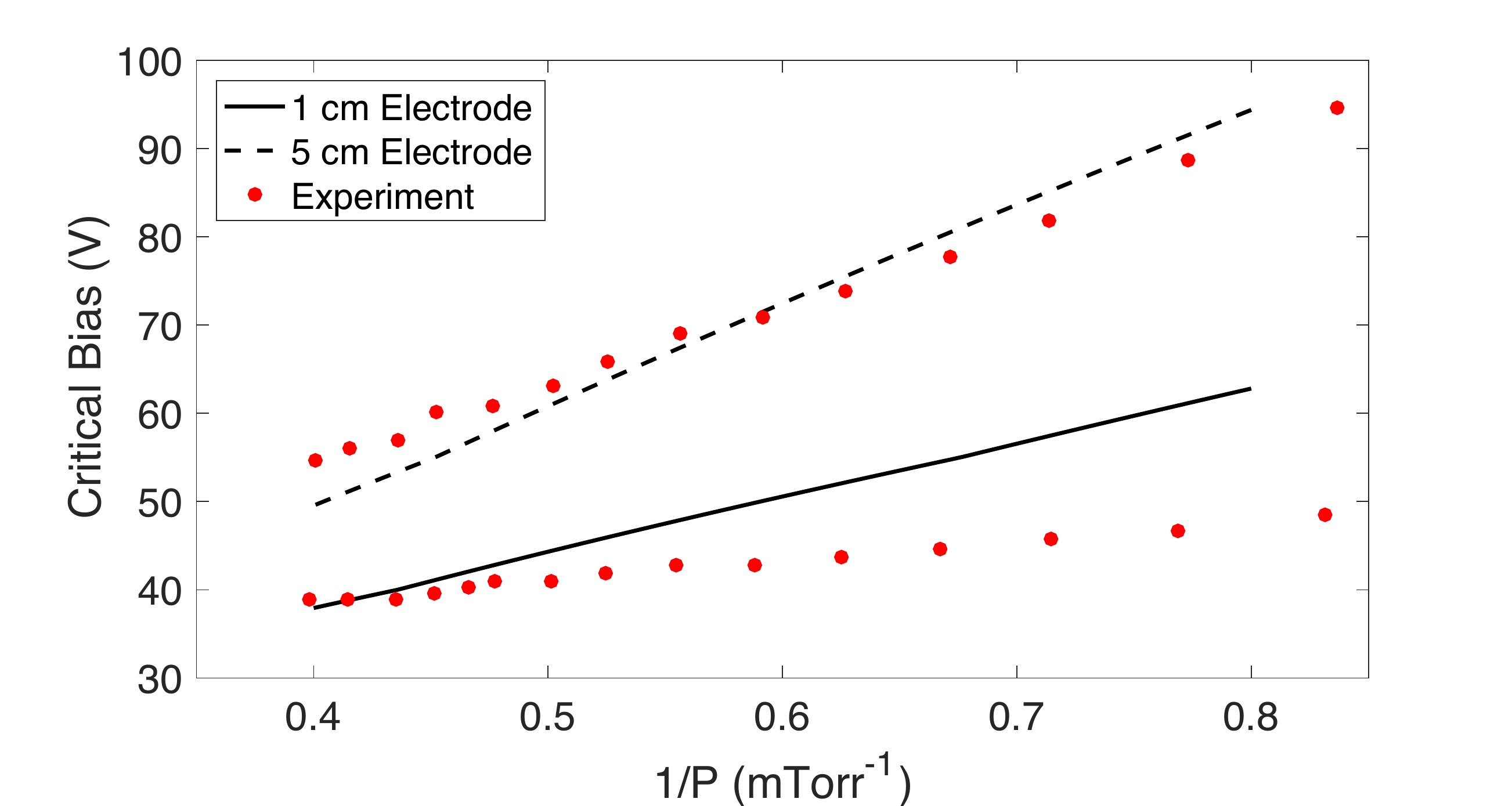}
\caption{The critical bias calculated from Eq. (\ref{tc2}) compared with experimental data for two different electrode diameters\cite{2009PSST...18c5002B}.\label{fg:cb}}
\end{figure}

An interesting observation from experiments is that increasing the size of the electrode decreases the critical bias\cite{2013PSST...22f5002Y,2009PSST...18c5002B}. In deriving Eq.~(\ref{tc}), it was assumed that the sheath and electrode have the same surface area. This is appropriate for the simulations of Sec.~\ref{sec:sim} where the initial approximately planar electron sheath is embedded within the ion sheath of the surrounding walls. However, for an isolated electrode, the thickness of the sheath increases its surface area. If the sheath surface area is modeled as a disk with diameter D and finite thickness $z_{\text{sheath}}$, the ratio of sheath area to electrode area is 
\begin{equation}
\frac{A_{\text{sheath}}}{A_{E}}=\frac{\pi D z_{\text{sheath}}+\pi(D/2)^2}{\pi(D/2)^2}.
\end{equation}
Using this, Eq. (\ref{tc}) is modified to include the sheath area that was unaccounted for,
\begin{equation}\label{tc2}
1-\frac{1}{2}\frac{A_{\text{sheath}}}{A_{E}}\sqrt{\frac{m_i}{m_e}}n_n\sigma_Iz_{\text{sheath}}=0.
\end{equation}
Figure~\ref{fg:cb} shows a solution of Eq.~(\ref{tc2}) for an argon plasma along with data from the experiments of Ref.~\onlinecite{2009PSST...18c5002B} conducted with 1 cm and 5.5 cm diameter electrodes. The plasma potential in the experiment was approximately 5~V, which was subtracted off the data points in Fig.~\ref{fg:cb}, although an independent measurement was not available at each pressure. The predicted values of the critical bias are found to be in good agreement with the experimental data. 

Sec.~\ref{sec:onsetsim} presented the results of a series of simulations used to determine the critical bias for a plasma with a 200 mTorr helium background, $T_e=2.4$ eV, and a bulk density of $3\times10^8 \ \textrm{cm}^{-3}$. The critical bias for these conditions was determined to be 41.6 V above the plasma potential. This is approximately 10 V more than the critical bias estimated from Fig.~\ref{fg:cb1}. One possibility for the agreement with $\sim1$~mTorr experiments in Fig.~\ref{fg:cb} and discrepancy for $\sim200$~mTorr simulations is that elastic collisions of electrons result in a thicker sheath than than that described by Eq. (\ref{esh})\cite{1991PhFlB...3.2796S}. A thicker sheath provides a greater loss area for ions formed by ionization, hence a greater ionization rate may be needed to satisfy Eq.~(\ref{tc2}).

\subsection{Anode spot current balance\label{sec:current}}

Like all steady-state plasmas, the anode spot maintains quasineutrality by the equal loss of electron and ion currents. Baalrud, Herskowitz, and Longmier developed a theory for the form of the sheath at a positive electrode using global current balance arguments\cite{2007PhPl...14d2109B}. In their work, the current loss to a positively biased electrode of area $A_E$ in a plasma chamber with wall area $A_W$ was considered. The form of the sheath was predicted to be either an electron sheath, ion sheath, or a sheath with a non-monotonic potential, depending on the ratio of surface area of the positively biased electrode to the wall area $A_E/A_W$. When the electrode area was small, an electron sheath was present since it would not significantly modify the balance of global current loss. However, when the electrode area was large the plasma potential locked to a few volts above the electrode potential preventing electrons from being lost at a faster rate than ions. This behavior has been observed in both experiment and simulation \cite{2014PhPl...21j3512B,2016PhPl...23f3519H}. 

\begin{figure}
\begin{center}
\includegraphics[scale=.5]{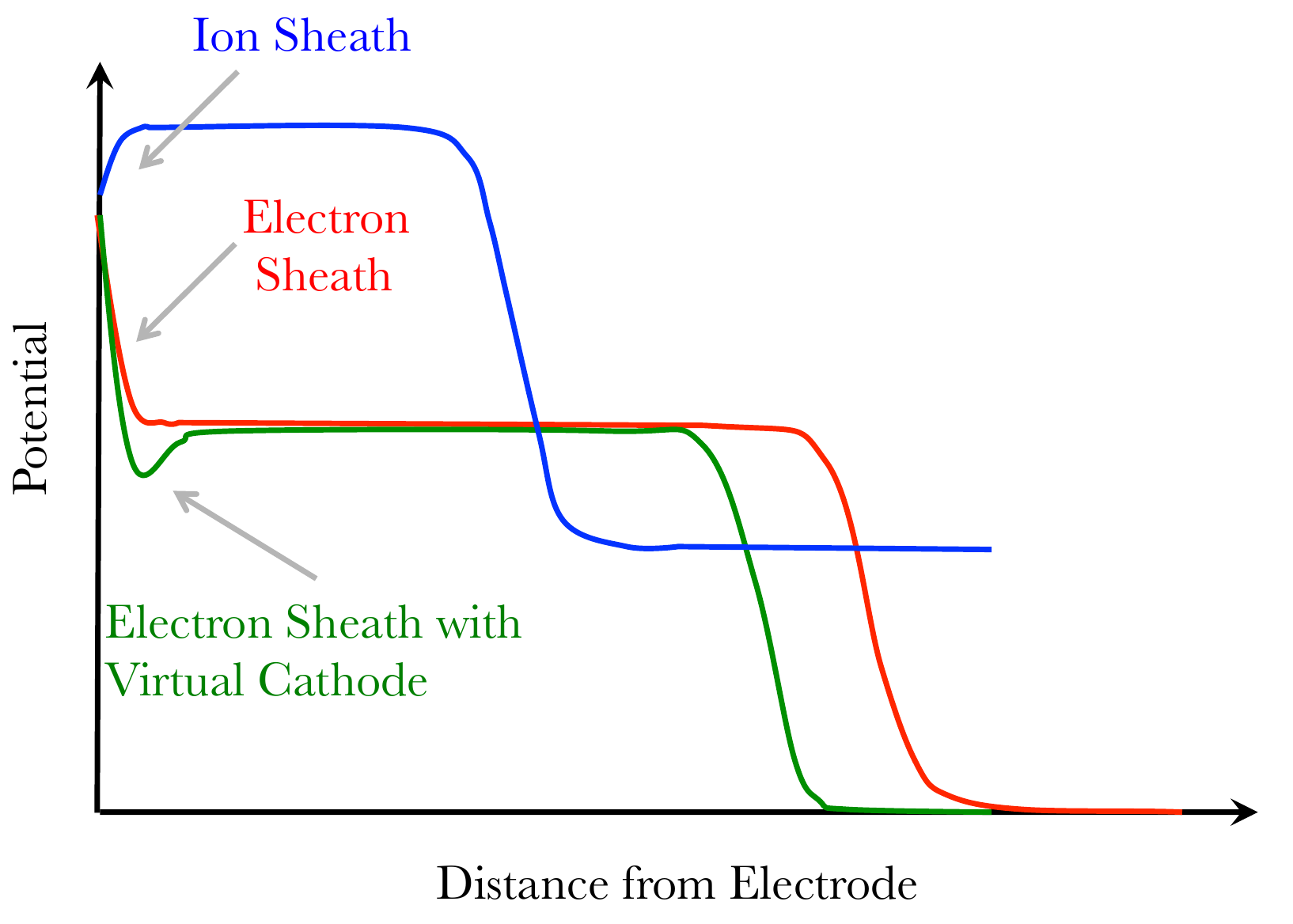}
\end{center}
\caption{Sketch of three types of potential structures associated with an anode spot. \label{fg:structure}
}
\end{figure}

Similar arguments can be applied to the anode spot by considering the current lost through a double layer of surface area $A_{\text{S}}$, which is the analog of $A_W$ for the quasineutral spot plasma, and the current lost to the electrode of area $A_E$. For this situation, the form of the sheath between the electrode and spot plasma is expected to be either an ion sheath, electron sheath, or a non-monotonic electron sheath with virtual cathode depending on the area ratio $A_{\text{S}}/A_E$. These possibilities are shown in Fig.~\ref{fg:structure}. Fig.~\ref{fg:t_profiles} shows that nearly all of the electrons in the spot are those formed by ionization of neutrals\footnote{This conclusion can verified by considering the high potential sheath edge density implied by the Langmuir condition along with the rarefaction of the bulk electron density by the strong double layer potential.}. Due to this observation, the analysis below will assume that the quasineutrailty condition within the anode spot is between the ions and $e^-_I$ electrons produced by ionization interactions in Eq. (\ref{ei1}) and (\ref{ei2}), i.e. $n_{eI}\approx n_i$.

\emph{Ion Sheath}: If the sheath between the spot plasma and electrode is an ion sheath, the electron current collected is $I_{eI}=e(n_{eI}\bar{v}_{T_{eI}}/4)A_E\exp(-e\Delta\phi_I/T_{eI})$, where $\Delta\phi_I=\phi_S-\phi_E$, $\phi_S$ is the anode spot plasma potential, and $\bar{v}_{T_{eI}}=\sqrt{8T_{eI}/\pi m_e}$ is the average speed of electrons born from ionization assuming a Maxwellian VDF with temperature $T_{eI}$. The ion current lost to the electrode and double layer is $I_i=0.6ec_sn_i(A_E+A_{\text{S}})$. Balancing the two currents gives the potential within the spot
\begin{equation}\label{AA}
-\frac{e\Delta\phi_I}{T_{eI}}=\ln\bigg[0.6\sqrt{2\pi\frac{T_{eS}}{T_{eI}}\frac{m_e}{m_i}}\bigg(\frac{A_{\text{S}}}{A_E}+1\bigg)\bigg].
\end{equation}
Here, $T_{eS}=(T_{eI}n_{eI}+T_{eB}n_{eB})/(n_{eI}+n_{eB})$ is the total electron temperature within the spot. Both $T_{eI}$ and $T_{eS}$ appear in this equation since the total electron temperature $T_{eS}$ determines the Bohm speed for ions and $T_{eI}$ determines the loss rate for ionization electrons. Although $n_{eI}\gg n_{eB}$, $T_{eI}\ll T_{eB}$ so both components must be considered in the total electron temperature $T_{eS}$. For Eq. (\ref{AA}) to be consistent with the assumption that an ion sheath is present ($\Delta\phi_I>0$), the area ratio must satisfy
\begin{equation}\label{a6}
\frac{A_{\text{S}}}{A_E}<0.6\sqrt{\frac{T_{eI}m_i}{2\pi T_{eS}m_e}}-1.
\end{equation}

\emph{Electron Sheath}: If the sheath between the spot plasma and electrode is an electron sheath, the electron current lost to the electrode is $I_{eI}=A_Een_{eI}v_{T_{eS}}$, where $v_{T_{eS}}=\sqrt{T_{eS}/m_e}$. 
The ion current lost to the electrode and double layer is $I_i=e(n_i\bar{v}_{T_i}/4)A_E\exp(-e\Delta\phi_E/T_i)+0.6en_ic_SA_{\text{S}}$, where $\Delta\phi_E=\phi_E-\phi_S$, and $\bar{v}_{T_i}=\sqrt{8T_{i}/\pi m_i}$ is the average speed for Maxwellian distributed ions with temperature $T_i$. The value of $\Delta\phi_E$ can be determined by balancing the two currents and using the quasineutrality condition resulting in 
\begin{equation}
\frac{e\Delta\phi_E}{T_i}=\ln\bigg[\sqrt{2\pi\frac{T_{eS}}{T_i}}\bigg(0.6\frac{A_{\text{S}}}{A_E}-\sqrt{\frac{m_i}{m_e}}\bigg)\bigg].
\end{equation}
For this to be consistent with the assumptions of an electron sheath at the electrode ($\Delta\phi_E>0$), 
\begin{equation}\label{a3}
\frac{A_{\text{S}}}{A_E}>0.6\bigg(\sqrt{\frac{T_i}{2\pi T_{eS}}}+\sqrt{\frac{m_i}{m_e}}\bigg).
\end{equation}

\emph{Electron Sheath with Virtual Cathode}: For area ratios between the electron and ion sheath cases the potential may have a non-monotonic profile such as a virtual cathode which limits the electron flux to the electrode\cite{2005PhPl...12e5502H}.

In experiments, larger anode spots at low pressure were observed to have electron sheath potential profiles both with and without virtual cathodes \cite{1991JPhD...24.1789S,2009PSST...18c5002B}. An exact comparison between these experiments and the theory of this section is not possible since the electron temperature was not reported. For smaller anode spots at higher pressures the potential profile is expected to include an ion sheath at the electrode. In the 2D simulation, the anode spot area is approximately half the circumference of a circle resulting in $A_{\text{S}}/A_E\approx3$. For this area ratio the sheath at the electrode is expected to be an ion sheath. This is consistent with the observed potential profile shown in Fig.~\ref{fg:simp}.


\subsection{Particle and power balance \& anode spot size\label{sec:power}}
In this subsection, the steady-state size of the anode spot is predicted by considering conservation of particle number and power in a global model of the spot plasma. This global model ties the potential across the double layer to the ionization rate within the spot, a feature that was not present in previous estimations of the anode spot size. Once the size and double layer potential are known, the form of the sheath at the electrode can be predicted using the theory of Sec.~\ref{sec:current}.  

\begin{figure} 
\begin{center}
\includegraphics[scale=.33]{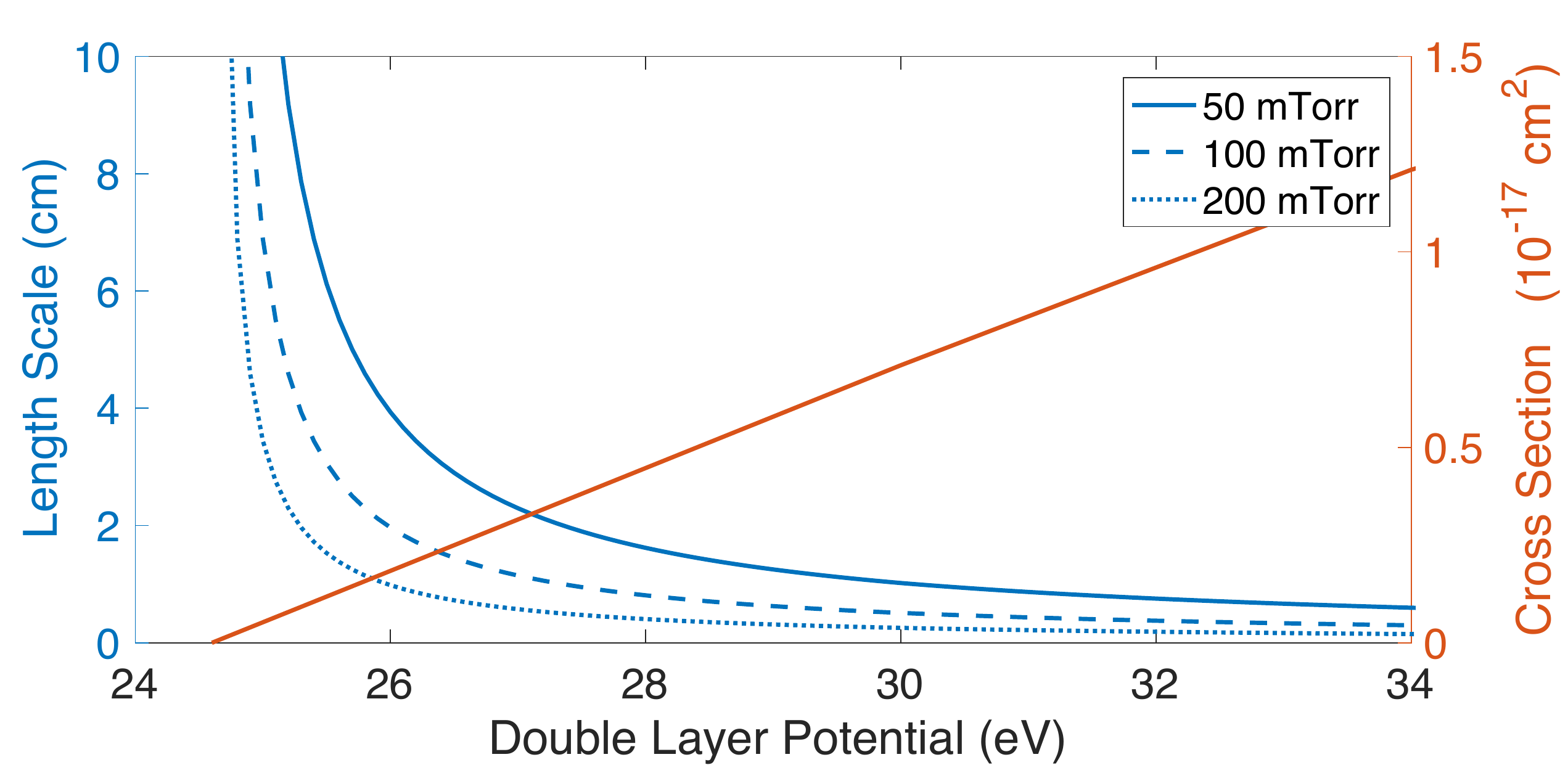}
\caption{The anode spot length scale L from Eq. (\ref{D}) as a function of the double layer potential for helium neutral pressures of 50, 100, and 200 mTorr. The electron impact ionization cross section obtained from LXcat\cite{lxcat} is also shown. \label{fg:length}}
\end{center}
\end{figure}

Balancing the volume ionization rate with the ion loss rate, and neglecting the loss of particles to the electrode\footnote{When the form of the sheath is an electron sheath or electron sheath with virtual cathode no ions are lost to the electrode, justifying this assumption. For the purposes of estimating the length scale for the ion sheath case,  $A_S$ can be considered the entire surface area bounding the spot plasma since the flux densities of ions directed at the electrode and double layer are the same. },
\begin{equation}\label{Bal1}
A_{S}n_sc_{s,s}=n_n\langle\sigma_I v\rangle_B n_B(\Delta\phi_{DL}) V_S.
\end{equation}
Here, $n_n$ is the neutral gas density, $\langle\sigma_I v\rangle_B$ is the rate constant for impact ionization by electrons accelerated from the bulk, $n_B(\Delta\phi_{DL}) $ is the density of the bulk electrons accelerated into the spot by $\Delta\phi_{DL}$, and $V_S$ is the anode spot volume. A length scale can be defined by relating the spot surface area and volume as $V_S=A_{\text{S}}L$. Eq. (\ref{Bal1}) can be written in terms of L,
\begin{equation}
L=\frac{n_S c_{s,S}}{n_n\langle\sigma_I v\rangle_B n_B(\Delta\phi_{DL})}.
\end{equation}
Using the approximation $\langle\sigma_I v\rangle_B n_B(\Delta\phi_{DL})\approx n_Bv_B\sigma_I$, assuming the cross section does not vary over the thermal width of the electron VDF, 
\
\begin{equation}\label{D}
L=\frac{n_S c_{s,S}}{n_n\sigma_I \Gamma_{e,B}}=\sqrt{\frac{m_e}{m_i}}\frac{1}{n_n\sigma_I},
\end{equation}
where the second equality results from the use of the Langmuir condition, returning the result of Eq. (\ref{ls}). The cross section in Eq. (\ref{D}) is strongly dependent on the energy of electrons gained while passing the double layer. The effect of this energy dependence was not previously considered. Instead, estimates based on a constant cross section were used to determine the length scale\cite{2009PSST...18c5002B,1991JPhD...24.1789S}. The length scale as a function of double layer energy is plotted from Eq. (\ref{D}) in Fig.~\ref{fg:length} for helium neutral background pressures of 50, 100, and 200 mTorr. To self consistently predict the size of the anode spot a constraint on the double layer potential is needed. 

In steady-state the power entering and leaving the spot plasma will also balance. Most of the bulk electrons entering the spot pass through without undergoing an ionizing collision and exit at the electrode. However, a small fraction ionize neutrals creating the spot plasma. For the non-ionizing electron component, the power entering and leaving the anode spot are in balance. The power input associated with the ionizing population is the number of ionization events per unit time multiplied by the residual energy left over after ionization of the neutral atom that goes into the kinetic energy of resulting particles. The power lost is the number of particles leaving the spot per unit time multiplied by the amount of energy that they carry with them. The resulting balance equation is
\begin{equation}\label{pw1}
\langle \sigma_I v \rangle_B n_n n_{e,B}V_S(e\Delta\phi_{DL}-\mathcal{E}_I)=\Gamma_i A_i \mathcal{E}_i+\Gamma_{eI}A_{eI}\mathcal{E}_{eI},
\end{equation}
where $\Gamma_{eI}$ and $\Gamma_i$ are the fluxes of $e^-_I$ electrons and ions leaving through their respective loss areas $A_{eI}$ and $A_i$ at energy $\mathcal{E}_{eI}$ and $\mathcal{E}_i$, and $\mathcal{E}_I$ is the threshold energy for electron impact ionization of neutrals.  Making the approximation $\langle \sigma_I v \rangle_B n_{e,B} \approx \Gamma_{e,B} \sigma_I$ by assuming the thermal width of electrons does not appreciably contribute to the ionization rate constant, Eq. (\ref{pw1}) is
\begin{equation}\label{pw2}
\Gamma_{e,B} \sigma_I n_n V_S(e\Delta\phi_{DL}-\mathcal{E}_I)=\Gamma_i A_i \mathcal{E}_i+\Gamma_{eI}A_{eI}\mathcal{E}_{eI}.
\end{equation}
This can be evaluated for each of the three different anode spot potential profile configurations discussed in Sec.~\ref{sec:current}. 

When the sheath at the anode is an electron sheath the loss area for ions is the double layer surface area $A_{\text{S}}$ and the loss area for electrons is $A_E$. Particle balance requires that the loss of electrons to the anode balance the loss of ions to the double layer resulting in 
\begin{equation}
n_Sv_{T_{eS}} A_E= n_S c_s A_{\text{S}}.
\end{equation}
Substituting this relation in the electron loss term of Eq. (\ref{pw2}) and using the length scale from Eq. (\ref{D}) along with $\mathcal{E}_i=T_{eS}/2$ for ions and $\mathcal{E}_{eI}=m_ev_{T_{eS}}^2/2=T_{eS}/2$ for electrons results in
\begin{equation} \label{pdl}
e\Delta\phi_{DL}=T_{eS}+\mathcal{E}_I.
\end{equation}

When the sheath at the anode is an ion sheath, ions are lost to the double layer and the electrode resulting in $A_i=A_{\text{S}}+A_E$. Once again, the ion flux is $\Gamma_i=n_Sc_s$ and the energy is  $\mathcal{E}_i=T_{eS}/2$. The electron loss only occurs to the anode since $\Delta\phi_{DL}$ is much greater than the difference in potential between the spot plasma and the anode. Particle balance results in $A_E \Gamma_{eI}=(A_{\text{S}}+A_E)\Gamma_i$. In this case, an electron lost to the anode is lost at the energy determined by the average speed of $e^-_I$ electrons, $\mathcal{E}_{eI}=m_e\vtei^2/2$. Using these considerations, the double layer potential is 
\begin{equation}\label{pdli}
e\Delta\phi_{DL}=\mathcal{E}_I+\bigg(\frac{T_{eS}}{2}+\frac{4T_{eI}}{\pi}\bigg)\bigg(\frac{A_E}{A_{\text{S}}}+1\bigg).
\end{equation}
Note that the use of different temperatures $T_{eI}$ and $T_{eS}$ is due to the total electron temperature $T_{eS}$ controlling the Bohm speed, while $T_{eI}$ controls the thermal flux of the particles of the dominant electron population in the spot ($e^-_I$ electrons).

Finally, consider the case where the sheath at the anode is an electron sheath with a virtual cathode, here the collection areas are the same as in the previous case. The flux of ions to the double layer is given by the same relation used in the electron sheath case, while the flux of electrons is reduced by the dip potential of the virtual cathode and is $\Gamma_{eI}=n_S\vtei e^{-e\Delta \phi_D/T_{eI}}$. Using particle balance along with $\mathcal{E}_i=T_{eS}/2$ for ions and $\mathcal{E}_{eI}=m_e\vtei^2/2$ the relation in Eq. (\ref{pw2}) reduces to 
\begin{equation}\label{pdld}
e\Delta\phi_{DL}=\mathcal{E}_I+\bigg(\frac{T_{eS}}{2}+\frac{4T_{eI}}{\pi}\bigg).
\end{equation}

The predictions of this section can be compared with the anode spot from the simulation in Sec.~\ref{sec:onsetsim}. Although these simulations never reached steady-state, the anode spot size varied slowly beginning around $10.7 \ \mu s$ after the initial double layer motion around $10.2 \ \mu s$, see the electric field magnitude in Fig.~\ref{fg:t_profiles}. Using the available particle velocity data from Fig.~\ref{fg:vdf}, the temperatures can be calculated allowing a comparison. See Sec. II of Ref. \onlinecite{2016PhPl...23h3510S} for details of the temperature calculation. From this data, the total electron temperature within the anode spot due to both $e^-_I$ and $e^-_B$ components is $T _{eS}=3.8$ eV, while the temperature for the $e_I^-$ component is $T_{eI}=1.15$ eV. At this time, the 2D electric field shows the area ratio of the double layer surface to electrode surface is roughly $A_{\text{S}}/A_E\approx3$. Inserting these quantities into Eq. (\ref{pdli}) results in $\Delta\phi_{DL}-\mathcal{E}_I/e\approx4.5$ V, which is close to the value $\Delta\phi_{DL}-\mathcal{E}_I/e\approx4.38$ V from the double layer potential in Fig.~\ref{fg:vdfloc}.

\subsection{Ion presheath in the anode spot\label{sec:presheath}}
At the high potential side of the double layer, ions leave the anode spot plasma entering into a positive space charge region. From the point of view of an ion within the spot plasma, this is similar to an ion entering an ion sheath. The Bohm criterion applies. An ion must enter the non-neutral region at a velocity exceeding it's sound speed. This velocity is attained in a presheath region leading up to the sheath edge. It was previously suggested that the anode spot presheath length was determined by the ion-neutral collision mean free path\cite{2009PSST...18c5002B}, although in some cases this would result in a presheath significantly longer than the observed size of the anode spot.  

The ion presheath length can be predicted by considering the rate constants for processes in the presheath. If the rate constant for ion-neutral collisions is greater than that for electron impact ionization, the presheath length would be determined by the ion-neutral mean free path. However, if the rate constant for electron impact ionization is larger, the presheath is dominated by ionization. In the latter case, the presheath length is predicted to be half of the plasma length, in this case half of the anode spot size L. This is one of the main results described by the various formulations of the Tonks-Langmuir presheath models\cite{1959PPS....74..145H,1980PhFl...23..803E}. In this section, the rate constants for these two processes is estimated, showing that the presheath is ionization dominated for typical plasma parameters.  

The rate constant is defined as $|\vc{v}_1-\vc{v}_2|\sigma(|\vc{v}_1-\vc{v}_2|)$ averaged over the distribution function of the incident and background particles denoted by subscript 1 and 2 respectively. This is 
\begin{equation}\label{K28}
K=\int d^3v_1 d^3v_2 f_1(\vc{v}_1)f_2(\vc{v}_2)\sigma(|\vc{v}_1-\vc{v}_2|)|\vc{v}_1-\vc{v}_2|,
\end{equation}
where $\sigma$ is the interaction cross section and the velocity distribution functions are normalized to unity. If the characteristic velocities of the background particles is much less than that of the incident particles then $|\vc{v}_1-\vc{v}_2|\approx |\vc{v}_1|$ and the $v_2$ integral can be evaluated.

 Eq. (\ref{K28}) is evaluated assuming the particles accelerated by the presheath have a flow shifted Maxwellian distribution, $f_1(\vc{v})=[1/(\pi^{3/2}v_T^3)]\exp[-(\vc{v}-\vc{U})^2/v_T^2]$, where $\vc{U}$ is the flow shift and $v_T=\sqrt{2T/m}$ is the thermal speed. Writing the integration variable in terms of energy $\mathcal{E}$ by using $v=\sqrt{2\mathcal{E}/m}$, and evaluating the integral in cylindrical coordinates aligned with the flow direction,

\begin{eqnarray}\label{K3}
&K(U,T)=\frac{4\pi}{\pi^{1/2}m}\int_0^\infty d\mathcal{E}\sqrt{\frac{\mathcal{E}}{T}}\sigma(\mathcal{E}) \nonumber\\
&  \times\exp\bigg[-\frac{\mathcal{E}}{T}-\bigg(\frac{U}{v_T}\bigg)^2\bigg]\frac{v_T}{U}\sinh\bigg(\frac{U}{v_T}\sqrt{\frac{\mathcal{E}}{T}}\bigg).
\end{eqnarray}

\begin{figure} 
\begin{center}
\includegraphics[scale=.35]{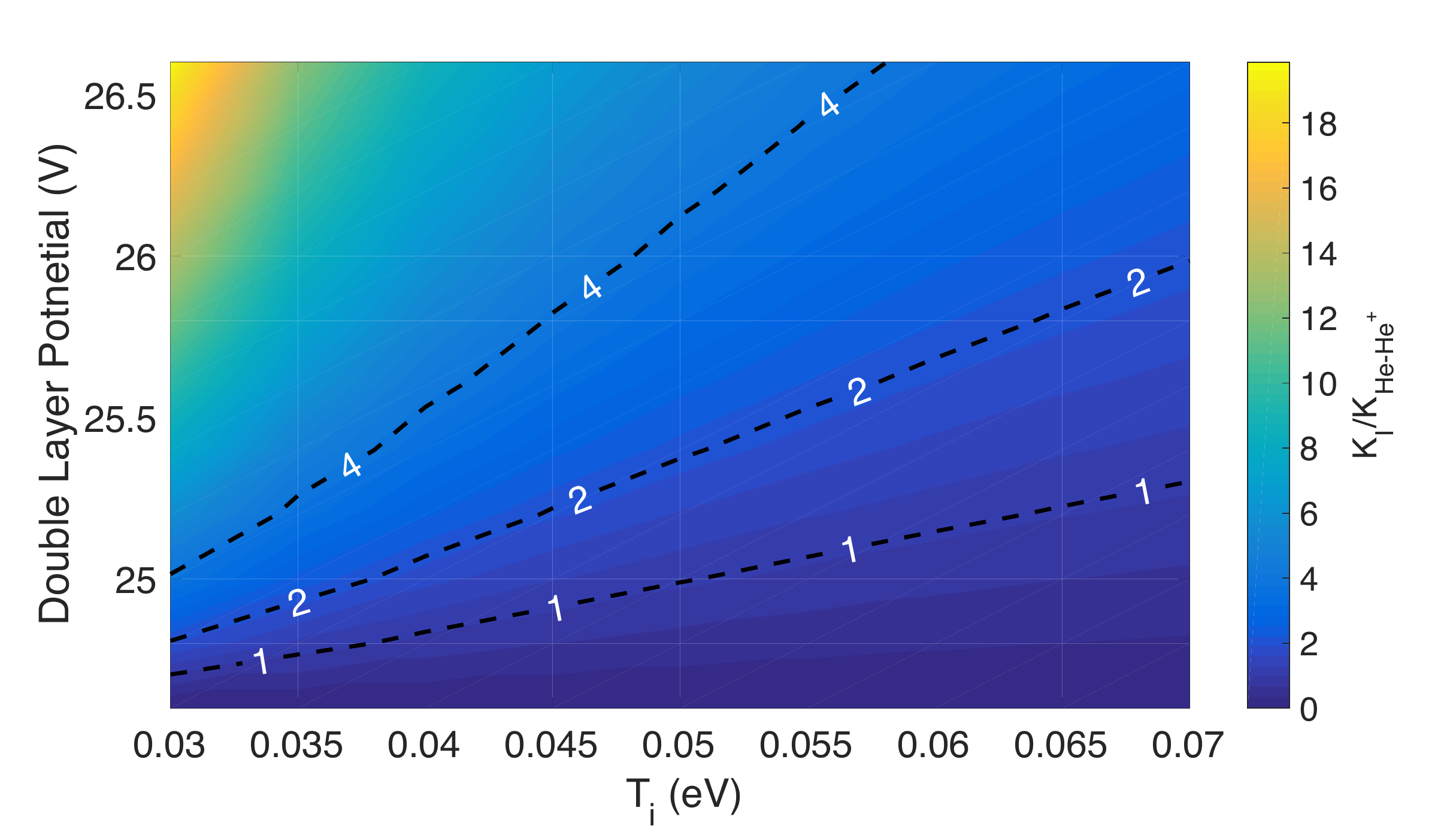}
\caption{The ratio of rate constants for electron impact ionization and elastic ion-neutral scattering in Helium as a function of ion temperature and double layer potential. Contours for  $K_I/K_{He^+-He}=$ 1, 2, and 4 are shown.\label{fg:ratio} }
\end{center}
\end{figure}

The ratio of the rate constants for electron impact ionization and elastic ion-neutral scattering  $K_I/K_{He^+-He}$ is considered to determine which processes is dominant. Because of the strong dependance on energy, the rate constant for ion neutral collisions is calculated using Eq. (\ref{K3}) with a flow of $c_s/2$ which is typical of a presheath ion. The ionization rate constant is estimated as
\begin{equation}
K_I\approx\sqrt{\frac{2 e\Delta\phi_{DL}}{m_e}}\sigma_I(e\Delta\phi_{DL}). 
\end{equation}
The ratio of these rate constants is shown in Fig.~\ref{fg:ratio}. The rate constant for electron impact ionization is several times that due to ion-neutral scattering for room temperature ions and values of $e\Delta\phi \sim  25eV$. Equations~(\ref{pdl})-(\ref{pdld}) predict that the double layer potential is approximately an electron temperature above the ionization energy of 24.6 eV. Since the temperatures encountered are typically $\gtrsim 1$ eV, this predicts that the presheath length will be determined by ionization processes for a helium neutral gas background. Similar results are obtained for argon plasmas.

\section{CONCLUSION\label{sec:conclusion}}

In this paper, the anode spot was studied for the first time using PIC simulations. These simulations demonstrated that electron impact ionization within the sheath results in a positive space charge layer adjacent to the electrode. With sufficient ionization, the positive space charge forms a potential well which traps low energy electrons formed by ionization in front of the electrode. This electron trapping results in an increase in electron density and formation of a quasineutral plasma. 

 A model for the spot onset was formulated based on observations of the simulated anode spot. The main feature of this model is that an imbalance in flux densities crossing the double layer leads to its motion and the expansion of the anode spot plasma. Using estimates of the sheath ionization rate, the value of the electrode bias relative to the plasma potential was tied to the sheath ionization. Predictions of the critical bias for spot onset were found to be in agreement with past experiments for different electrode sizes and plasma conditions, allowing an experimental test of the spot onset model. 

Steady-state properties were predicted based on an analysis of current, power, and particle balance of the spot plasma. Maintenance of quasineutrality within the spot dictates the form of the sheath between the anode spot plasma and the electrode, determining how particles are lost from the anode spot. In the model, balance of the total ionization rate and particle loss rate determines the anode spot size as a function of the energy of electrons entering the spot from the bulk. The size is determined once the double layer potential is known. Balance of power lost from and deposited into the spot plasma sets this potential in the model. The predicted energy gain of an electron crossing the double layer potential is predicted to be a few electron volts above the ionization energy for typical experimental conditions. This result is consistent with several anode spot experiments.

\section*{Acknowledgments}
The authors thank Nathaniel Shaffer for his comments on the manuscript. This research was supported by the Office of Fusion Energy Science at the U.S. Department of Energy under contract DE-AC04-94SL85000. The first author was also supported by the U.S. Department of Energy, Office of Science, Office of Workforce Development for Teachers and Scientists, Office of Science Graduate Student Research (SCGSR) program. The SCGSR program is administered by the Oak Ridge Institute for Science and Education for the DOE under contract number DE-AC05-06OR23100.

\bibliography{spotV2}

\begin{thebibliography}{36}%
\makeatletter
\providecommand \@ifxundefined [1]{%
 \@ifx{#1\undefined}
}%
\providecommand \@ifnum [1]{%
 \ifnum #1\expandafter \@firstoftwo
 \else \expandafter \@secondoftwo
 \fi
}%
\providecommand \@ifx [1]{%
 \ifx #1\expandafter \@firstoftwo
 \else \expandafter \@secondoftwo
 \fi
}%
\providecommand \natexlab [1]{#1}%
\providecommand \enquote  [1]{``#1''}%
\providecommand \bibnamefont  [1]{#1}%
\providecommand \bibfnamefont [1]{#1}%
\providecommand \citenamefont [1]{#1}%
\providecommand \href@noop [0]{\@secondoftwo}%
\providecommand \href [0]{\begingroup \@sanitize@url \@href}%
\providecommand \@href[1]{\@@startlink{#1}\@@href}%
\providecommand \@@href[1]{\endgroup#1\@@endlink}%
\providecommand \@sanitize@url [0]{\catcode `\\12\catcode `\$12\catcode
  `\&12\catcode `\#12\catcode `\^12\catcode `\_12\catcode `\%12\relax}%
\providecommand \@@startlink[1]{}%
\providecommand \@@endlink[0]{}%
\providecommand \url  [0]{\begingroup\@sanitize@url \@url }%
\providecommand \@url [1]{\endgroup\@href {#1}{\urlprefix }}%
\providecommand \urlprefix  [0]{URL }%
\providecommand \Eprint [0]{\href }%
\providecommand \doibase [0]{http://dx.doi.org/}%
\providecommand \selectlanguage [0]{\@gobble}%
\providecommand \bibinfo  [0]{\@secondoftwo}%
\providecommand \bibfield  [0]{\@secondoftwo}%
\providecommand \translation [1]{[#1]}%
\providecommand \BibitemOpen [0]{}%
\providecommand \bibitemStop [0]{}%
\providecommand \bibitemNoStop [0]{.\EOS\space}%
\providecommand \EOS [0]{\spacefactor3000\relax}%
\providecommand \BibitemShut  [1]{\csname bibitem#1\endcsname}%
\let\auto@bib@innerbib\@empty
\bibitem [{Note1()}]{Note1}%
  \BibitemOpen
  \bibinfo {note} {Note that the term {\protect \emph {double layer}} usually
  refers to a similar structure between a high and low potential plasma with
  equal amounts of positive and negative charge, hence the need to distinguish
  anode double layers.}\BibitemShut {Stop}%
\bibitem [{\citenamefont {{Baalrud}}, \citenamefont {{Longmier}},\ and\
  \citenamefont {{Hershkowitz}}(2009)}]{2009PSST...18c5002B}%
  \BibitemOpen
  \bibfield  {author} {\bibinfo {author} {\bibfnamefont {S.~D.}\ \bibnamefont
  {{Baalrud}}}, \bibinfo {author} {\bibfnamefont {B.}~\bibnamefont
  {{Longmier}}}, \ and\ \bibinfo {author} {\bibfnamefont {N.}~\bibnamefont
  {{Hershkowitz}}},\ }\href {\doibase 10.1088/0963-0252/18/3/035002} {\bibfield
   {journal} {\bibinfo  {journal} {Plasma Sources Science Technology}\ }\textbf
  {\bibinfo {volume} {18}},\ \bibinfo {eid} {035002} (\bibinfo {year}
  {2009})}\BibitemShut {NoStop}%
\bibitem [{\citenamefont {{Langmuir}}(1929)}]{1929PhRv...33..954L}%
  \BibitemOpen
  \bibfield  {author} {\bibinfo {author} {\bibfnamefont {I.}~\bibnamefont
  {{Langmuir}}},\ }\href {\doibase 10.1103/PhysRev.33.954} {\bibfield
  {journal} {\bibinfo  {journal} {Physical Review}\ }\textbf {\bibinfo {volume}
  {33}},\ \bibinfo {pages} {954} (\bibinfo {year} {1929})}\BibitemShut
  {NoStop}%
\bibitem [{\citenamefont {{Torven}}\ and\ \citenamefont
  {{Andersson}}(1979)}]{1979JPhD...12..717T}%
  \BibitemOpen
  \bibfield  {author} {\bibinfo {author} {\bibfnamefont {S.}~\bibnamefont
  {{Torven}}}\ and\ \bibinfo {author} {\bibfnamefont {D.}~\bibnamefont
  {{Andersson}}},\ }\href {\doibase 10.1088/0022-3727/12/5/012} {\bibfield
  {journal} {\bibinfo  {journal} {Journal of Physics D Applied Physics}\
  }\textbf {\bibinfo {volume} {12}},\ \bibinfo {pages} {717} (\bibinfo {year}
  {1979})}\BibitemShut {NoStop}%
\bibitem [{\citenamefont {{Stenzel}}, \citenamefont {{Ionita}},\ and\
  \citenamefont {{Schrittwieser}}(2008)}]{2008PSST...17c5006S}%
  \BibitemOpen
  \bibfield  {author} {\bibinfo {author} {\bibfnamefont {R.~L.}\ \bibnamefont
  {{Stenzel}}}, \bibinfo {author} {\bibfnamefont {C.}~\bibnamefont {{Ionita}}},
  \ and\ \bibinfo {author} {\bibfnamefont {R.}~\bibnamefont
  {{Schrittwieser}}},\ }\href {\doibase 10.1088/0963-0252/17/3/035006}
  {\bibfield  {journal} {\bibinfo  {journal} {Plasma Sources Science
  Technology}\ }\textbf {\bibinfo {volume} {17}},\ \bibinfo {eid} {035006}
  (\bibinfo {year} {2008})}\BibitemShut {NoStop}%
\bibitem [{\citenamefont {{Yip}}\ \emph {et~al.}(2013)\citenamefont {{Yip}},
  \citenamefont {{Sheehan}}, \citenamefont {{Hershkowitz}},\ and\ \citenamefont
  {{Severn}}}]{2013PSST...22f5002Y}%
  \BibitemOpen
  \bibfield  {author} {\bibinfo {author} {\bibfnamefont {C.-S.}\ \bibnamefont
  {{Yip}}}, \bibinfo {author} {\bibfnamefont {J.~P.}\ \bibnamefont
  {{Sheehan}}}, \bibinfo {author} {\bibfnamefont {N.}~\bibnamefont
  {{Hershkowitz}}}, \ and\ \bibinfo {author} {\bibfnamefont {G.}~\bibnamefont
  {{Severn}}},\ }\href {\doibase 10.1088/0963-0252/22/6/065002} {\bibfield
  {journal} {\bibinfo  {journal} {Plasma Sources Science Technology}\ }\textbf
  {\bibinfo {volume} {22}},\ \bibinfo {eid} {065002} (\bibinfo {year}
  {2013})}\BibitemShut {NoStop}%
\bibitem [{\citenamefont {{Barkan}}\ and\ \citenamefont
  {{Merlino}}(1995)}]{1995PhPl....2.3261B}%
  \BibitemOpen
  \bibfield  {author} {\bibinfo {author} {\bibfnamefont {A.}~\bibnamefont
  {{Barkan}}}\ and\ \bibinfo {author} {\bibfnamefont {R.~L.}\ \bibnamefont
  {{Merlino}}},\ }\href {\doibase 10.1063/1.871159} {\bibfield  {journal}
  {\bibinfo  {journal} {Physics of Plasmas}\ }\textbf {\bibinfo {volume} {2}},\
  \bibinfo {pages} {3261} (\bibinfo {year} {1995})}\BibitemShut {NoStop}%
\bibitem [{\citenamefont {{Ahedo}}(1996)}]{1996PhPl....3.3875A}%
  \BibitemOpen
  \bibfield  {author} {\bibinfo {author} {\bibfnamefont {E.}~\bibnamefont
  {{Ahedo}}},\ }\href {\doibase 10.1063/1.871575} {\bibfield  {journal}
  {\bibinfo  {journal} {Physics of Plasmas}\ }\textbf {\bibinfo {volume} {3}},\
  \bibinfo {pages} {3875} (\bibinfo {year} {1996})}\BibitemShut {NoStop}%
\bibitem [{\citenamefont {{Park}}\ \emph {et~al.}(2011)\citenamefont {{Park}},
  \citenamefont {{Lee}}, \citenamefont {{Chung}},\ and\ \citenamefont
  {{Hwang}}}]{2011RScI...82l3303P}%
  \BibitemOpen
  \bibfield  {author} {\bibinfo {author} {\bibfnamefont {Y.-S.}\ \bibnamefont
  {{Park}}}, \bibinfo {author} {\bibfnamefont {Y.}~\bibnamefont {{Lee}}},
  \bibinfo {author} {\bibfnamefont {K.-J.}\ \bibnamefont {{Chung}}}, \ and\
  \bibinfo {author} {\bibfnamefont {Y.~S.}\ \bibnamefont {{Hwang}}},\ }\href
  {\doibase 10.1063/1.3664616} {\bibfield  {journal} {\bibinfo  {journal}
  {Review of Scientific Instruments}\ }\textbf {\bibinfo {volume} {82}},\
  \bibinfo {pages} {123303} (\bibinfo {year} {2011})}\BibitemShut {NoStop}%
\bibitem [{\citenamefont {{Sanduloviciu}}(2013)}]{2013JMPh....4..364S}%
  \BibitemOpen
  \bibfield  {author} {\bibinfo {author} {\bibfnamefont {M.}~\bibnamefont
  {{Sanduloviciu}}},\ }\href {\doibase 10.4236/jmp.2013.43051} {\bibfield
  {journal} {\bibinfo  {journal} {Journal of Modern Physics}\ }\textbf
  {\bibinfo {volume} {4}},\ \bibinfo {pages} {364} (\bibinfo {year}
  {2013})}\BibitemShut {NoStop}%
\bibitem [{\citenamefont {{Maszl}}, \citenamefont {{Laimer}},\ and\
  \citenamefont {{Stori}}(2011)}]{2011ITPS...39.2118M}%
  \BibitemOpen
  \bibfield  {author} {\bibinfo {author} {\bibfnamefont {C.}~\bibnamefont
  {{Maszl}}}, \bibinfo {author} {\bibfnamefont {J.}~\bibnamefont {{Laimer}}}, \
  and\ \bibinfo {author} {\bibfnamefont {H.}~\bibnamefont {{Stori}}},\ }\href
  {\doibase 10.1109/TPS.2011.2157365} {\bibfield  {journal} {\bibinfo
  {journal} {IEEE Transactions on Plasma Science}\ }\textbf {\bibinfo {volume}
  {39}},\ \bibinfo {pages} {2118} (\bibinfo {year} {2011})}\BibitemShut
  {NoStop}%
\bibitem [{\citenamefont {{Block}}(1978)}]{1978Ap&SS..55...59B}%
  \BibitemOpen
  \bibfield  {author} {\bibinfo {author} {\bibfnamefont {L.~P.}\ \bibnamefont
  {{Block}}},\ }\href {\doibase 10.1007/BF00642580} {\bibfield  {journal}
  {\bibinfo  {journal} {Astrophysics and Space Science}\ }\textbf {\bibinfo
  {volume} {55}},\ \bibinfo {pages} {59} (\bibinfo {year} {1978})}\BibitemShut
  {NoStop}%
\bibitem [{\citenamefont {{Block}}(1972)}]{1972CosEl...3..349B}%
  \BibitemOpen
  \bibfield  {author} {\bibinfo {author} {\bibfnamefont {L.~P.}\ \bibnamefont
  {{Block}}},\ }\href@noop {} {\bibfield  {journal} {\bibinfo  {journal}
  {Cosmic Electrodynamics}\ }\textbf {\bibinfo {volume} {3}},\ \bibinfo {pages}
  {349} (\bibinfo {year} {1972})}\BibitemShut {NoStop}%
\bibitem [{\citenamefont {{Song}}, \citenamefont {{Merlino}},\ and\
  \citenamefont {{D'Angelo}}(1992)}]{1992PhyS...45..391S}%
  \BibitemOpen
  \bibfield  {author} {\bibinfo {author} {\bibfnamefont {B.}~\bibnamefont
  {{Song}}}, \bibinfo {author} {\bibfnamefont {R.~L.}\ \bibnamefont
  {{Merlino}}}, \ and\ \bibinfo {author} {\bibfnamefont {N.}~\bibnamefont
  {{D'Angelo}}},\ }\href {\doibase 10.1088/0031-8949/45/4/018} {\bibfield
  {journal} {\bibinfo  {journal} {Physica Scripta}\ }\textbf {\bibinfo {volume}
  {45}},\ \bibinfo {pages} {391} (\bibinfo {year} {1992})}\BibitemShut
  {NoStop}%
\bibitem [{\citenamefont {{Song}}, \citenamefont {{D'Angelo}},\ and\
  \citenamefont {{Merlino}}(1992)}]{1992JPhD...25..938S}%
  \BibitemOpen
  \bibfield  {author} {\bibinfo {author} {\bibfnamefont {B.}~\bibnamefont
  {{Song}}}, \bibinfo {author} {\bibfnamefont {N.}~\bibnamefont {{D'Angelo}}},
  \ and\ \bibinfo {author} {\bibfnamefont {R.~L.}\ \bibnamefont {{Merlino}}},\
  }\href {\doibase 10.1088/0022-3727/25/6/006} {\bibfield  {journal} {\bibinfo
  {journal} {Journal of Physics D Applied Physics}\ }\textbf {\bibinfo {volume}
  {25}},\ \bibinfo {pages} {938} (\bibinfo {year} {1992})}\BibitemShut
  {NoStop}%
\bibitem [{\citenamefont {{Andersson}}(1981)}]{1981JPhD...14.1403A}%
  \BibitemOpen
  \bibfield  {author} {\bibinfo {author} {\bibfnamefont {D.}~\bibnamefont
  {{Andersson}}},\ }\href {\doibase 10.1088/0022-3727/14/8/008} {\bibfield
  {journal} {\bibinfo  {journal} {Journal of Physics D Applied Physics}\
  }\textbf {\bibinfo {volume} {14}},\ \bibinfo {pages} {1403} (\bibinfo {year}
  {1981})}\BibitemShut {NoStop}%
\bibitem [{\citenamefont {{Song}}, \citenamefont {{D'Angelo}},\ and\
  \citenamefont {{Merlino}}(1991)}]{1991JPhD...24.1789S}%
  \BibitemOpen
  \bibfield  {author} {\bibinfo {author} {\bibfnamefont {B.}~\bibnamefont
  {{Song}}}, \bibinfo {author} {\bibfnamefont {N.}~\bibnamefont {{D'Angelo}}},
  \ and\ \bibinfo {author} {\bibfnamefont {R.~L.}\ \bibnamefont {{Merlino}}},\
  }\href {\doibase 10.1088/0022-3727/24/10/012} {\bibfield  {journal} {\bibinfo
   {journal} {Journal of Physics D Applied Physics}\ }\textbf {\bibinfo
  {volume} {24}},\ \bibinfo {pages} {1789} (\bibinfo {year}
  {1991})}\BibitemShut {NoStop}%
\bibitem [{\citenamefont {{Timko}}\ \emph {et~al.}(2012)\citenamefont
  {{Timko}}, \citenamefont {{Crozier}}, \citenamefont {{Hopkins}},
  \citenamefont {{Matyash}},\ and\ \citenamefont
  {{Schneider}}}]{2012CoPP...52..295T}%
  \BibitemOpen
  \bibfield  {author} {\bibinfo {author} {\bibfnamefont {H.}~\bibnamefont
  {{Timko}}}, \bibinfo {author} {\bibfnamefont {P.~S.}\ \bibnamefont
  {{Crozier}}}, \bibinfo {author} {\bibfnamefont {M.~M.}\ \bibnamefont
  {{Hopkins}}}, \bibinfo {author} {\bibfnamefont {K.}~\bibnamefont
  {{Matyash}}}, \ and\ \bibinfo {author} {\bibfnamefont {R.}~\bibnamefont
  {{Schneider}}},\ }\href {\doibase 10.1002/ctpp.201100051} {\bibfield
  {journal} {\bibinfo  {journal} {Contributions to Plasma Physics}\ }\textbf
  {\bibinfo {volume} {52}},\ \bibinfo {pages} {295} (\bibinfo {year}
  {2012})}\BibitemShut {NoStop}%
\bibitem [{\citenamefont {Bird}(1998)}]{bird1998molecular}%
  \BibitemOpen
  \bibfield  {author} {\bibinfo {author} {\bibfnamefont {G.}~\bibnamefont
  {Bird}},\ }\href {https://books.google.com/books?id=xd2knQEACAAJ} {\emph
  {\bibinfo {title} {Molecular Gas Dynamics and the Direct Simulation of Gas
  Flows}}},\ Oxford engineering science series\ (\bibinfo  {publisher}
  {Clarendon Press},\ \bibinfo {year} {1998})\BibitemShut {NoStop}%
\bibitem [{\citenamefont {{Barnat}}, \citenamefont {{Laity}},\ and\
  \citenamefont {{Baalrud}}(2014)}]{2014PhPl...21j3512B}%
  \BibitemOpen
  \bibfield  {author} {\bibinfo {author} {\bibfnamefont {E.~V.}\ \bibnamefont
  {{Barnat}}}, \bibinfo {author} {\bibfnamefont {G.~R.}\ \bibnamefont
  {{Laity}}}, \ and\ \bibinfo {author} {\bibfnamefont {S.~D.}\ \bibnamefont
  {{Baalrud}}},\ }\href {\doibase 10.1063/1.4897927} {\bibfield  {journal}
  {\bibinfo  {journal} {Physics of Plasmas}\ }\textbf {\bibinfo {volume}
  {21}},\ \bibinfo {eid} {103512} (\bibinfo {year} {2014})}\BibitemShut
  {NoStop}%
\bibitem [{lxc()}]{lxcat}%
  \BibitemOpen
  \href@noop {} {\enquote {\bibinfo {title} {Phelps database, www.lxcat.net,
  retrieved on september 30, 2015.}}\ }\BibitemShut {NoStop}%
\bibitem [{\citenamefont {{Scheiner}}\ \emph {et~al.}(2015)\citenamefont
  {{Scheiner}}, \citenamefont {{Baalrud}}, \citenamefont {{Yee}}, \citenamefont
  {{Hopkins}},\ and\ \citenamefont {{Barnat}}}]{2015PhPl...22l3520S}%
  \BibitemOpen
  \bibfield  {author} {\bibinfo {author} {\bibfnamefont {B.}~\bibnamefont
  {{Scheiner}}}, \bibinfo {author} {\bibfnamefont {S.~D.}\ \bibnamefont
  {{Baalrud}}}, \bibinfo {author} {\bibfnamefont {B.~T.}\ \bibnamefont
  {{Yee}}}, \bibinfo {author} {\bibfnamefont {M.~M.}\ \bibnamefont
  {{Hopkins}}}, \ and\ \bibinfo {author} {\bibfnamefont {E.~V.}\ \bibnamefont
  {{Barnat}}},\ }\href {\doibase 10.1063/1.4939024} {\bibfield  {journal}
  {\bibinfo  {journal} {Physics of Plasmas}\ }\textbf {\bibinfo {volume}
  {22}},\ \bibinfo {eid} {123520} (\bibinfo {year} {2015})},\ \Eprint
  {http://arxiv.org/abs/1510.03490} {arXiv:1510.03490 [physics.plasm-ph]}
  \BibitemShut {NoStop}%
\bibitem [{\citenamefont {{Cartier}}\ and\ \citenamefont
  {{Merlino}}(1987)}]{1987PhFl...30.2549C}%
  \BibitemOpen
  \bibfield  {author} {\bibinfo {author} {\bibfnamefont {S.~L.}\ \bibnamefont
  {{Cartier}}}\ and\ \bibinfo {author} {\bibfnamefont {R.~L.}\ \bibnamefont
  {{Merlino}}},\ }\href {\doibase 10.1063/1.866093} {\bibfield  {journal}
  {\bibinfo  {journal} {Physics of Fluids}\ }\textbf {\bibinfo {volume} {30}},\
  \bibinfo {pages} {2549} (\bibinfo {year} {1987})}\BibitemShut {NoStop}%
\bibitem [{\citenamefont {{Baalrud}}, \citenamefont {{Hershkowitz}},\ and\
  \citenamefont {{Longmier}}(2007)}]{2007PhPl...14d2109B}%
  \BibitemOpen
  \bibfield  {author} {\bibinfo {author} {\bibfnamefont {S.~D.}\ \bibnamefont
  {{Baalrud}}}, \bibinfo {author} {\bibfnamefont {N.}~\bibnamefont
  {{Hershkowitz}}}, \ and\ \bibinfo {author} {\bibfnamefont {B.}~\bibnamefont
  {{Longmier}}},\ }\href {\doibase 10.1063/1.2722262} {\bibfield  {journal}
  {\bibinfo  {journal} {Physics of Plasmas}\ }\textbf {\bibinfo {volume}
  {14}},\ \bibinfo {eid} {042109} (\bibinfo {year} {2007})}\BibitemShut
  {NoStop}%
\bibitem [{\citenamefont {{Scheiner}}\ \emph {et~al.}(2016)\citenamefont
  {{Scheiner}}, \citenamefont {{Baalrud}}, \citenamefont {{Hopkins}},
  \citenamefont {{Yee}},\ and\ \citenamefont {{Barnat}}}]{2016PhPl...23h3510S}%
  \BibitemOpen
  \bibfield  {author} {\bibinfo {author} {\bibfnamefont {B.}~\bibnamefont
  {{Scheiner}}}, \bibinfo {author} {\bibfnamefont {S.~D.}\ \bibnamefont
  {{Baalrud}}}, \bibinfo {author} {\bibfnamefont {M.~M.}\ \bibnamefont
  {{Hopkins}}}, \bibinfo {author} {\bibfnamefont {B.~T.}\ \bibnamefont
  {{Yee}}}, \ and\ \bibinfo {author} {\bibfnamefont {E.~V.}\ \bibnamefont
  {{Barnat}}},\ }\href {\doibase 10.1063/1.4960382} {\bibfield  {journal}
  {\bibinfo  {journal} {Physics of Plasmas}\ }\textbf {\bibinfo {volume}
  {23}},\ \bibinfo {eid} {083510} (\bibinfo {year} {2016})},\ \Eprint
  {http://arxiv.org/abs/1604.08251} {arXiv:1604.08251 [physics.plasm-ph]}
  \BibitemShut {NoStop}%
\bibitem [{\citenamefont {{Yee}}\ \emph {et~al.}(2017)\citenamefont {{Yee}},
  \citenamefont {{Scheiner}}, \citenamefont {{Baalrud}}, \citenamefont
  {{Barnat}},\ and\ \citenamefont {{Hopkins}}}]{2017PSST...26b5009Y}%
  \BibitemOpen
  \bibfield  {author} {\bibinfo {author} {\bibfnamefont {B.~T.}\ \bibnamefont
  {{Yee}}}, \bibinfo {author} {\bibfnamefont {B.}~\bibnamefont {{Scheiner}}},
  \bibinfo {author} {\bibfnamefont {S.~D.}\ \bibnamefont {{Baalrud}}}, \bibinfo
  {author} {\bibfnamefont {E.~V.}\ \bibnamefont {{Barnat}}}, \ and\ \bibinfo
  {author} {\bibfnamefont {M.~M.}\ \bibnamefont {{Hopkins}}},\ }\href {\doibase
  10.1088/1361-6595/aa56d7} {\bibfield  {journal} {\bibinfo  {journal} {Plasma
  Sources Science Technology}\ }\textbf {\bibinfo {volume} {26}},\ \bibinfo
  {eid} {025009} (\bibinfo {year} {2017})}\BibitemShut {NoStop}%
\bibitem [{Note2()}]{Note2}%
  \BibitemOpen
  \bibinfo {note} {Here, onset is defined as the expansion of the quasineutral
  region. This coincides with the motion of the double layer.}\BibitemShut
  {Stop}%
\bibitem [{\citenamefont {{Conde}}, \citenamefont {{Ferro Font{'a}n}},\ and\
  \citenamefont {{Lamb{'a}s}}(2006)}]{2006PhPl...13k3504C}%
  \BibitemOpen
  \bibfield  {author} {\bibinfo {author} {\bibfnamefont {L.}~\bibnamefont
  {{Conde}}}, \bibinfo {author} {\bibfnamefont {C.}~\bibnamefont {{Ferro
  Font{'a}n}}}, \ and\ \bibinfo {author} {\bibfnamefont {J.}~\bibnamefont
  {{Lamb{'a}s}}},\ }\href {\doibase 10.1063/1.2388265} {\bibfield  {journal}
  {\bibinfo  {journal} {Physics of Plasmas}\ }\textbf {\bibinfo {volume}
  {13}},\ \bibinfo {eid} {113504} (\bibinfo {year} {2006})}\BibitemShut
  {NoStop}%
\bibitem [{\citenamefont {{Andersson}}\ and\ \citenamefont
  {{Sorensen}}(1983)}]{1983JPhD...16..601A}%
  \BibitemOpen
  \bibfield  {author} {\bibinfo {author} {\bibfnamefont {D.}~\bibnamefont
  {{Andersson}}}\ and\ \bibinfo {author} {\bibfnamefont {J.}~\bibnamefont
  {{Sorensen}}},\ }\href {\doibase 10.1088/0022-3727/16/4/020} {\bibfield
  {journal} {\bibinfo  {journal} {Journal of Physics D Applied Physics}\
  }\textbf {\bibinfo {volume} {16}},\ \bibinfo {pages} {601} (\bibinfo {year}
  {1983})}\BibitemShut {NoStop}%
\bibitem [{\citenamefont {{Sheridan}}\ and\ \citenamefont
  {{Goree}}(1991)}]{1991PhFlB...3.2796S}%
  \BibitemOpen
  \bibfield  {author} {\bibinfo {author} {\bibfnamefont {T.~E.}\ \bibnamefont
  {{Sheridan}}}\ and\ \bibinfo {author} {\bibfnamefont {J.}~\bibnamefont
  {{Goree}}},\ }\href {\doibase 10.1063/1.859987} {\bibfield  {journal}
  {\bibinfo  {journal} {Physics of Fluids B}\ }\textbf {\bibinfo {volume}
  {3}},\ \bibinfo {pages} {2796} (\bibinfo {year} {1991})}\BibitemShut
  {NoStop}%
\bibitem [{\citenamefont {{Hopkins}}\ \emph {et~al.}(2016)\citenamefont
  {{Hopkins}}, \citenamefont {{Yee}}, \citenamefont {{Baalrud}},\ and\
  \citenamefont {{Barnat}}}]{2016PhPl...23f3519H}%
  \BibitemOpen
  \bibfield  {author} {\bibinfo {author} {\bibfnamefont {M.~M.}\ \bibnamefont
  {{Hopkins}}}, \bibinfo {author} {\bibfnamefont {B.~T.}\ \bibnamefont
  {{Yee}}}, \bibinfo {author} {\bibfnamefont {S.~D.}\ \bibnamefont
  {{Baalrud}}}, \ and\ \bibinfo {author} {\bibfnamefont {E.~V.}\ \bibnamefont
  {{Barnat}}},\ }\href {\doibase 10.1063/1.4953896} {\bibfield  {journal}
  {\bibinfo  {journal} {Physics of Plasmas}\ }\textbf {\bibinfo {volume}
  {23}},\ \bibinfo {eid} {063519} (\bibinfo {year} {2016})}\BibitemShut
  {NoStop}%
\bibitem [{Note3()}]{Note3}%
  \BibitemOpen
  \bibinfo {note} {This conclusion can verified by considering the high
  potential sheath edge density implied by the Langmuir condition along with
  the rarefaction of the bulk electron density by the strong double layer
  potential.}\BibitemShut {Stop}%
\bibitem [{\citenamefont {{Hershkowitz}}(2005)}]{2005PhPl...12e5502H}%
  \BibitemOpen
  \bibfield  {author} {\bibinfo {author} {\bibfnamefont {N.}~\bibnamefont
  {{Hershkowitz}}},\ }\href {\doibase 10.1063/1.1887189} {\bibfield  {journal}
  {\bibinfo  {journal} {Physics of Plasmas}\ }\textbf {\bibinfo {volume}
  {12}},\ \bibinfo {eid} {055502} (\bibinfo {year} {2005})}\BibitemShut
  {NoStop}%
\bibitem [{Note4()}]{Note4}%
  \BibitemOpen
  \bibinfo {note} {When the form of the sheath is an electron sheath or
  electron sheath with virtual cathode no ions are lost to the electrode,
  justifying this assumption. For the purposes of estimating the length scale
  for the ion sheath case, $A_S$ can be considered the entire surface area
  bounding the spot plasma since the flux densities of ions directed at the
  electrode and double layer are the same.}\BibitemShut {Stop}%
\bibitem [{\citenamefont {{Harrison}}\ and\ \citenamefont
  {{Thompson}}(1959)}]{1959PPS....74..145H}%
  \BibitemOpen
  \bibfield  {author} {\bibinfo {author} {\bibfnamefont {E.~R.}\ \bibnamefont
  {{Harrison}}}\ and\ \bibinfo {author} {\bibfnamefont {W.~B.}\ \bibnamefont
  {{Thompson}}},\ }\href {\doibase 10.1088/0370-1328/74/2/301} {\bibfield
  {journal} {\bibinfo  {journal} {Proceedings of the Physical Society}\
  }\textbf {\bibinfo {volume} {74}},\ \bibinfo {pages} {145} (\bibinfo {year}
  {1959})}\BibitemShut {NoStop}%
\bibitem [{\citenamefont {{Emmert}}\ \emph {et~al.}(1980)\citenamefont
  {{Emmert}}, \citenamefont {{Wieland}}, \citenamefont {{Mense}},\ and\
  \citenamefont {{Davidson}}}]{1980PhFl...23..803E}%
  \BibitemOpen
  \bibfield  {author} {\bibinfo {author} {\bibfnamefont {G.~A.}\ \bibnamefont
  {{Emmert}}}, \bibinfo {author} {\bibfnamefont {R.~M.}\ \bibnamefont
  {{Wieland}}}, \bibinfo {author} {\bibfnamefont {A.~T.}\ \bibnamefont
  {{Mense}}}, \ and\ \bibinfo {author} {\bibfnamefont {J.~N.}\ \bibnamefont
  {{Davidson}}},\ }\href {\doibase 10.1063/1.863062} {\bibfield  {journal}
  {\bibinfo  {journal} {Physics of Fluids}\ }\textbf {\bibinfo {volume} {23}},\
  \bibinfo {pages} {803} (\bibinfo {year} {1980})}\BibitemShut {NoStop}%
\end{thebibliography}%

\end{document}